\documentclass{aa}  

\usepackage{graphicx}
\usepackage{txfonts}
\usepackage[table]{xcolor}
\usepackage{hyperref}
\usepackage{booktabs}
\usepackage{comment}
\usepackage{enumitem}
\usepackage{subcaption}
\usepackage{multirow}

\usepackage{diagbox}  

\newcommand{\JWST}{\textit{JWST }}
\newcommand{\HST}{\textit{HST }}
\newcommand{\re}{r_\mathrm{eff}}

\newcommand{\galfit}{\textsc{galfit }}
\newcommand{\statmorph}{\textsc{statmorph }}
\newcommand{\sersic}{S\'ersic }

\begin{document} 

    \title{Galaxy morphologies at cosmic noon with \textit{JWST}: A foundation for exploring gas transport with bars and spiral arms}
    \titlerunning{Galaxy morphologies at cosmic noon with \JWST}

   \author{
    Juan M. Espejo Salcedo\inst{1}\thanks{\email{jespejo@mpe.mpg.de}}
    \and
    Stavros Pastras\inst{1,2}
    \and
    Josef Vácha\inst{3}
    \and
    Claudia Pulsoni\inst{1}
    \and
    Reinhard Genzel\inst{1}
    \and
    N. M. Förster Schreiber\inst{1}
    \and
    Jean-Baptiste Jolly\inst{1}
    \and
    Capucine Barfety\inst{1}
    \and
    Jianhang Chen\inst{1}
    \and
    Giulia Tozzi\inst{1}
    \and
    Daizhong Liu\inst{1,4}
    \and
    Lilian L. Lee\inst{1}
    \and
    Stijn Wuyts\inst{5}
    \and
    Linda J. Tacconi\inst{1}
    \and
    Ric Davies\inst{1}
    \and
    Hannah Übler\inst{1}
    \and
    Dieter Lutz\inst{1}
    \and
    Emily Wisnioski\inst{6,7}
    \and
    Jinyi Shangguan\inst{1,8}
    \and
    Minju Lee\inst{9,10}
    \and
    Sedona H. Price\inst{11}
    \and
    Frank Eisenhauer\inst{1}
    \and
    Alvio Renzini\inst{12}
    \and
    Amit Nestor Shachar\inst{13}
    \and
    Rodrigo Herrera-Camus\inst{14}
}

\institute{
    Max-Planck-Institut für extraterrestische Physik (MPE), Giessenbachstraße 1., 85748 Garching, Germany
    \and
    Max-Planck-Institut für Astrophysik (MPA), Karl-Schwarzschild-Str. 1, 85748, Garching, Germany
\and
    Department of Physics, University of Oxford, Clarendon Laboratory, Parks Road, OX1 3PU Oxford, UK
    \and
    Purple Mountain Observatory, Chinese Academy of Sciences, 10 Yuanhua Road, Nanjing 210023, China
    \and
    Department of Physics, University of Bath, Claverton Down, Bath BA2 7AY, UK
    \and
    Research School of Astronomy and Astrophysics, Australian National University, Weston Creek, ACT 2611, Australia
    \and
    ARC Centre of Excellence for All-Sky Astrophysics in 3 Dimensions (ASTRO 3D), Canberra, ACT 2611, Australia
    \and
    Kavli Institute for Astronomy and Astrophysics, Peking University, Beijing 100871, People’s Republic of China
    \and
    Cosmic Dawn Center (DAWN), Denmark
    \and
    DTU-Space, Technical University of Denmark, Elektrovej 327, DK2800 Kgs. Lyngby, Denmark
    \and
    Department of Physics and Astronomy and PITT PACC, University of Pittsburgh, Pittsburgh, PA 15260, USA
    \and
    Osservatorio Astronomico di Padova, Vicolo dell’Osservatorio 5, Padova I-35122, Italy
    \and
    School of Physics and Astronomy, Tel Aviv University, Tel Aviv 69978, Israel
    \and
    Universidad de Concepci\'on, Víctor Lamas 1290, Barrio Universitario, Concepci\'on, Chile
}

   \date{Received 24 March 2025 / Accepted 11 June 2025}



\abstract{The way in which radial flows shape galaxy structure and evolution remains an open question. Internal drivers of such flows, such as bars and spiral arms, known to mediate gas flows in the local Universe, are now observable at high redshift thanks to \textit{JWST}'s unobscured view. We investigated the morphology of massive star-forming galaxies at $0.8<z<1.3$ and $2.0<z<2.5$, epochs marking the peak and decline of cosmic star formation, both well covered by kinematic surveys. Using \textit{JWST}/NIRCam imaging, we visually classified 1,451 galaxies, identified non-axisymmetric features, counted the number of spiral arms, analyzed nonparametric morphological indicators, and studied the dynamical support of the sample covered by kinematics ($\approx11\%$ of the sample) as measured via $v/\sigma$. Disk galaxies dominate the sample (fraction $0.82\pm0.03$); among them, $0.48\pm0.04$ exhibit spiral structure and $0.11\pm0.03$ host bars. Both fractions decline with redshift, in agreement with previous studies. The proportion of two- and three-armed spirals remains largely unchanged across our redshift bins: approximately two-thirds show two arms and one-third show three arms in both bins. Notably, we find a higher incidence of three-armed spirals ($\approx 0.30$) than reported in the local Universe ($\approx 0.20$), suggesting a mild evolution in spiral arm multiplicity. Nonparametric morphological metrics strongly correlate with stellar mass but show no significant redshift evolution. Finally, kinematic analysis reveals a correlation between disk morphology and rotational support: most disks exhibit $v/\sigma>3$ and median values of $v/\sigma>7$ for spirals and $v/\sigma>5$ for barred galaxies. This study establishes a population-wide framework for linking galaxy morphology and dynamics at cosmic noon, providing a key reference for future studies on the role of detailed structural features in galaxy evolution.}

   \keywords{galaxies: disks -- galaxies: morphology -- galaxies: evolution}

   \maketitle

\section{Introduction}
\label{section: Introduction}

A fundamental question in galaxy evolution is how early star-forming galaxies assembled the well-ordered structures seen in the present-day Universe. While early morphological studies suggested that high-redshift galaxies were highly irregular and dynamically unstable, kinematic surveys have since revealed that disk-like rotation is widespread at cosmic noon ($1<z<3$) (e.g., \citealt{Forster_2008_no_AO, Law_2, KMOS3D, Stott_2016, Simons_2017, Harrison_2017, Uebler_2019}). Integral field spectroscopy has consistently shown that a large fraction of star-forming galaxies at these epochs exhibit strong rotational support, often with velocity dispersions lower than expected for turbulent irregular morphologies. These findings challenged the prevailing view based on \textit{HST} imaging, which primarily identified galaxies with clumpy chaotic structures (e.g., \citealp{Abraham_1996, Conselice_2000, Jogee_2004, Lotz_2006, Elmegreen_2007, Sheth_2008}).

With the advent of the \textit{HST} Wide Field Camera 3 (WFC3) and near-infrared imaging, deeper surveys began to reveal a growing number of galaxies with more regular morphologies. By tracing rest-frame optical emission, these observations identified an increasing fraction of disk-like systems and even reported the presence of spiral arms and bars in some of them—though the full prevalence of disks and their substructures remained uncertain due to resolution and surface brightness limitations (e.g., \citealp{Wuyts_2011, Van_der_Wel_2014, Morishita_2014, Guo_2015, Kartaltepe_2015}).

The arrival of \textit{JWST} has dramatically transformed our ability to study galaxy structure at high redshift, offering a combination of longer wavelengths, increased depth, and unprecedented resolution. Recent studies leveraging \textit{JWST} NIRCam imaging consistently reveal a high fraction of disk galaxies at and beyond cosmic noon (e.g., \citealt{Jacobs_2023, Ferreira_2023, Kartaltepe_2023, Liu_2023, Nelson_2023, Huertas-Company_2024, Lee_2024_JWST_morphologies, Tohill_2024, Pandya_2024, Smethurst_2025}). Additionally, some studies provide clear evidence of spiral arms and bars (e.g., \citealt{Constantin_2023, Ferreira_2023, Guo_2023, Kuhn_2024, LeConte_2024, Mengyuan_2025, Chugunov_2025, Geron_2025}). Thus, many galaxies that once appeared irregular or clumpy in \textit{HST} imaging are now recognized as well-ordered disks, some of which exhibit prominent non-axisymmetric features.

Beyond their prevalence, these nonsymmetric morphological features are key agents of secular evolution, regulating gas flows and star formation within galaxies. In the local Universe, bars and spiral arms are known to redistribute angular momentum and funnel gas toward galaxy centers, often fueling central star formation and bulge growth. As demonstrated by \cite{Yu_2022} with the EDGE-CALIFA survey of local galaxies, spiral arms and bars both efficiently channel gas inward, leading to higher central gas concentrations and elevated star formation rates.

If these mechanisms were already at play at cosmic noon, they could have profound implications for the early buildup of galaxy structure. Recent kinematic studies have uncovered evidence for rapid radial inflows in high-redshift main-sequence disks (e.g., \citealp{Genzel_2023}, \citealp{Pastras_2025}), which suggests that these processes may have been even more pronounced in the past. These findings contradict earlier expectations that non-axisymmetric structures such as spiral arms and bars would be short-lived in gas-rich turbulent environments (e.g., \citealt{Genzel_2008, Ceverino_2010, Bournaud_2011}). Instead, they suggest that such features were not only prevalent, but they also played a fundamental role in shaping galaxy evolution throughout cosmic time. Given these advancements, establishing a direct connection between morphology and kinematics is crucial for understanding how high-redshift disks evolve and whether non-axisymmetric features are the primary drivers of internal gas flows and structural transformation.

For this study we exploited \JWST NIRCam imaging to investigate the morphologies of massive main-sequence galaxies at $1<z<3$; we focused on two key redshift intervals that trace distinct phases of cosmic star formation activity: a peak period at $2.0<z<2.5$ and a later decline at $0.8<z<1.3$ (e.g., \citealt{Madau_SFH, Forster_review_2020}). These bins also match redshift ranges for which H$\alpha$ is observable from the ground, and large H$\alpha$ kinematics samples are available.

Our goal is to establish a robust morphological foundation for interpreting the growing body of kinematic measurements from near-infrared integral field spectroscopy and millimeter and/or micromillimeter observations. Specifically, our sample selection aligns with that of major spectroscopic surveys such as KMOS$^{\rm 3D}$, SINS, KROSS, NOEMA-3D, and PHIBSS, thus enabling the connections between the morphological and kinematic analyses. By providing a comprehensive visual classification of galaxy structure (including disks, bars, and spiral arms), this work serves as a key reference for future studies whose aim is to investigate the connections between morphology, gas transport, and secular evolution. Additionally, we complement our visual classifications with quantitative (nonparametric) morphological metrics, leveraging well-established statistical methods to characterize galaxy structure and to systematically link morphology and dynamics at cosmic noon.

In Sect. \ref{section: Sample}, we describe the sample selection, followed by a discussion of the imaging data, its corresponding data reduction methods, and the assembly of the kinematic sample used in this study. Section \ref{section: Galaxy morphology} presents the overall morphological properties of the sample, detailing the visual classification scheme, general classification results, and the identification of bars and spiral structures. Section \ref{section: Morphological evolution} examines the evolution of these morphological features as a function of redshift and compares our findings with previous studies. In Sect. \ref{section: Morphological metrics}, we perform a quantitative analysis of galaxy morphology, exploring nonparametric structural indicators and their relationship to the visual classifications. In Sect. \ref{section: Kinematics} we investigate the connection between morphology and kinematics, and discuss the implications for galaxy evolution. Finally, Sect. \ref{section: Conclusions} summarizes the key findings of this work. Additional details, including discussions of observational limitations, preparations for the morphological analysis, and a description of the methodology used to separate galaxy morphologies based on structural metrics are provided in Appendices \ref{appendix - Observational limitations}-\ref{appendix: Separation Efficiency}.

We adopt a $\Lambda$CDM cosmology with $\Omega_\mathrm{m}$ = 0.3, $\Omega_\Lambda$ = 0.7, and $H_0$ = 70 km s$^{-1}$ Mpc$^{-1}$. In this framework, one arcsecond corresponds to $8.1$ kpc at $z=1.05$ and $8.24$ kpc at $z=2.25$, which are the mid-points of the two chosen redshift bins.

\begin{figure}
	\includegraphics[width=0.47\textwidth]{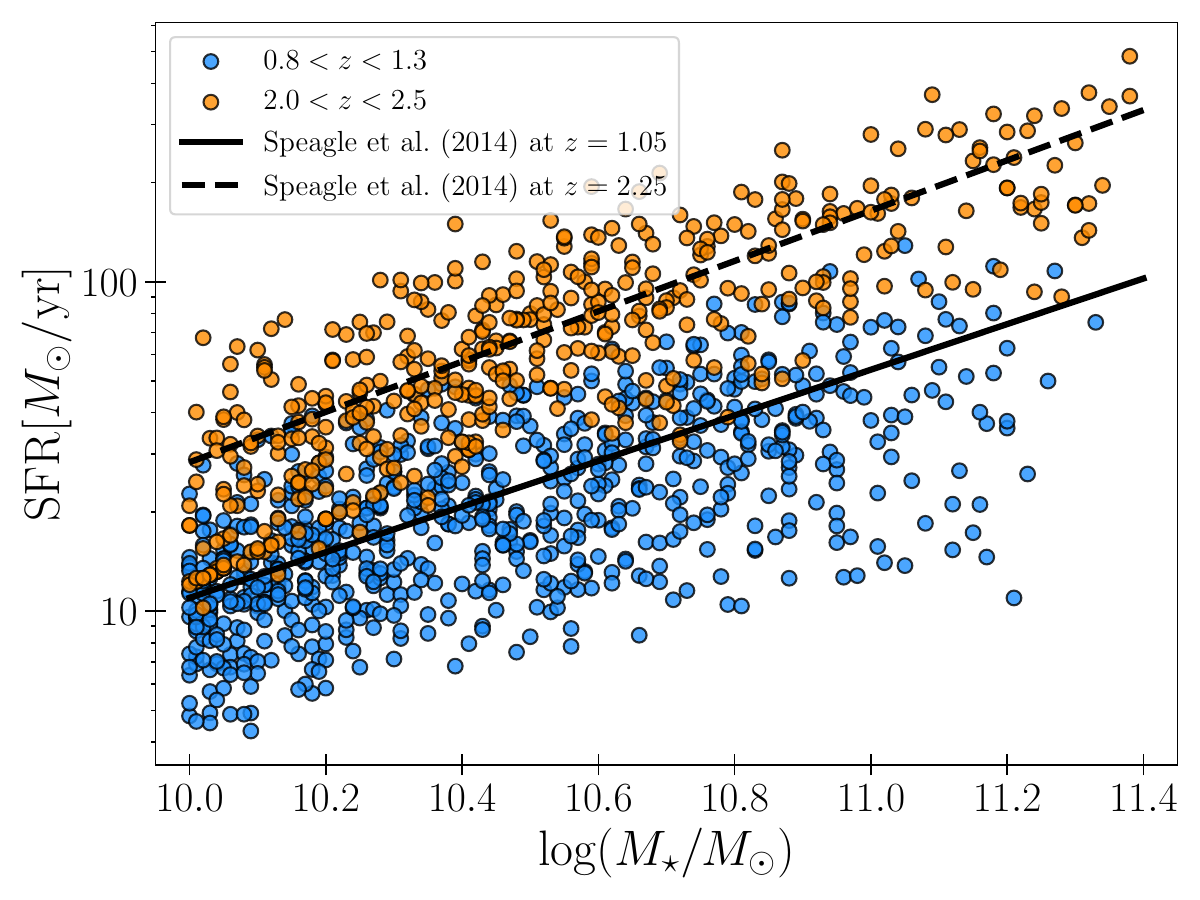}
	\caption{Star formation rate (SFR) \rm{vs} stellar mass $M_\star$ for the selected sample. The lines represent the \protect\cite{Speagle_2014} relations for the mid-points of the redshift bins $z=1.05$ and $z=2.25$.}
    \label{fig: Main sequence}
\end{figure}

\section{Sample}
\label{section: Sample}

\begin{figure}
	\includegraphics[width=0.49\textwidth]{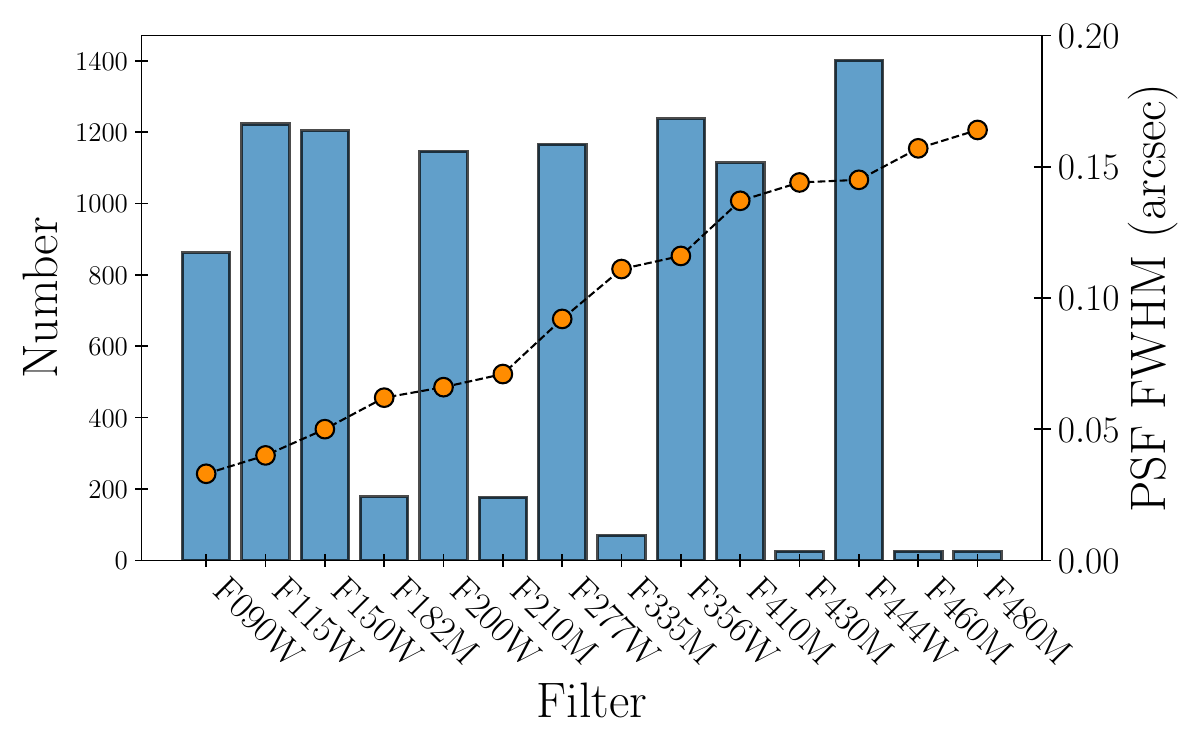}
	\caption{Number of galaxies observed in each filter. The most frequently observed filter combination—F090W, F115W, F150W, F200W, F277W, F356W, F410M, and F444W—is shared by 739 galaxies (51\% of the sample), meaning these galaxies were observed in all these filters. The orange dots indicate the empirical PSF FWHM in each filter. To reduce overcrowding, we omitted F250M and F300M, which contain only ten galaxies each.}
    \label{fig: Galaxies per filter}
\end{figure}

\subsection{NIRCam imaging}
\label{subsection: NIRCam imaging}

To investigate morphological evolution at cosmic noon, we focus on two epochs corresponding to $0.8<z<1.3$ and $2.0<z<2.5$. Since star-forming galaxies dominate the population at these epochs, we apply selection criteria targeting massive star-forming galaxies with star formation rates (SFR) within 0.4 dex of the star formation main sequence (MS), as parameterized by \cite{Whitaker2014}, i.e., $\log (\mathrm{SFR}/\mathrm{SFR_{MS}}) = [-0.4,0.4]$. Additionally, we choose masses within $\log(M_\star/M_{\odot}) = [10.0, 11.4]$, and sizes in the range $\re \, \mathrm{(arcsec)} = [0.2,1.0]$. The criteria are designed to represent the bulk of typical star-forming galaxies at high masses.

These quantities are derived from the 3D-HST catalog by \cite{Momcheva_3DHST} (see also \citealp{Brammer_3DHST}; \citealp{Skelton_2014}), which serves as the parent sample for this study. The 3D-HST survey combines near-infrared grism spectroscopy ($1.1$–$1.7,\mu$m, $R \sim 130$ plus parallel $0.5-0.9 \mu m$ grism observations) with deep multi-wavelength photometry from the five CANDELS fields, yielding high-quality redshifts and rest-frame properties for over 100,000 galaxies. At redshifts $z \sim 1-2.5$, the sample is broadly representative of the star-forming galaxy population with $\log(M_\star/M_\odot) \gtrsim 9$, and redshift precision of $\sigma_z/(1+z) \lesssim 0.003$ for bright sources. Its dynamic range in stellar mass and SFR makes it an ideal parent sample for statistical studies of galaxy structure and evolution at cosmic noon. Global galaxy properties such as stellar mass and star formation rates are derived following the prescriptions by \cite{Wuyts_2011_ladder_of_SFRs}. Figure \ref{fig: Main sequence} provides a visualization of the sample distribution in the SFR \rm{vs} $M_\star$ plane across the two redshift bins.

Building on this 3D-HST-based selection, we collected available NIRCam imaging from public data releases of major \JWST surveys, specifically JADES \citep{Finkelstein_2023} and CEERS \citep{Bagley_2023}. These data span the five CANDELS fields (\cite{Koekemoer_2011}; \cite{Grogin_2011}): the UKIDSS Ultra Deep Survey (UDS; \citealt{Lawrence_2007, Cirasuolo_2007}), the Extended Groth Strip (EGS; \citealt{Davies_2007}), the Cosmic Evolution Survey (COSMOS; \citealt{Scoville_2007, Koekemoer_2007}), and the Great Observatories Origins Deep Survey (GOODS) North and South (\citealt{Giavalisco_2004}). Table \ref{tab: Sample coverage} provides an overview of the sample sizes in each of the CANDELS fields.

For each galaxy, we use a uniform pixel scale of 0.025 arcsec per pixel and retrieve the NIRCam images across all available filters covering the range \(0.9 \, \mu \mathrm{m} < \lambda < 4.8 \, \mu \mathrm{m}\). Among the used filters, seven are broadband (F090W, F115W, F150W, F200W, F277W, F356W, and F444W), and nine are mediumband (F182M, F210M, F250M, F300M, F335M, F410M, F430M, F460M, and F480M). The most commonly available filter combination in our dataset consists of eight filters—F090W, F115W, F150W, F200W, F277W, F356W, F410M, and F444W— and covers 739 galaxies, making it particularly useful for assessing filter dependence in morphological measurements. Our sample consists of a total of 1,451 galaxies (9,877 NIRCam images in total), where 811 and 640 are in the low- and high-redshift bins, respectively. Figure \ref{fig: Galaxies per filter} presents the number of galaxies observed in each filter along with the corresponding empirical PSF size in arcseconds\footnote{Empirical PSF sizes used in this work are taken from \url{https://jwst-docs.stsci.edu/jwst-near-infrared-camera}}. As shown by \cite{Lee_2024_JWST_morphologies}, using empirically derived PSFs from field stars in different deep fields has a negligible impact on the inferred morphological fractions across all fields and introduces only minor systematic biases in \sersic indices.

\begin{table}
    \caption{Distribution of the number of galaxies over the CANDELS deep fields and redshift bins.}
    \label{tab: Sample coverage}
    \centering
\begin{tabular}{cccc}
\hline\hline
Field  & Number   & $0.8<z<1.3$   & $2.0<z<2.5$ \\
\hline
UDS            & 399   & 199 & 200 \\
EGS            & 298   & 165 & 133 \\
COSMOS         & 399   & 243 & 156 \\
GOODS-South    & 155   & 83  & 72  \\
GOODS-North    & 200   & 121 & 79  \\
All fields     & 1,451 & 811 & 640 \\
\bottomrule
\end{tabular}
\end{table}

\begin{figure*}
\centering
	\includegraphics[width=0.99
    \textwidth]{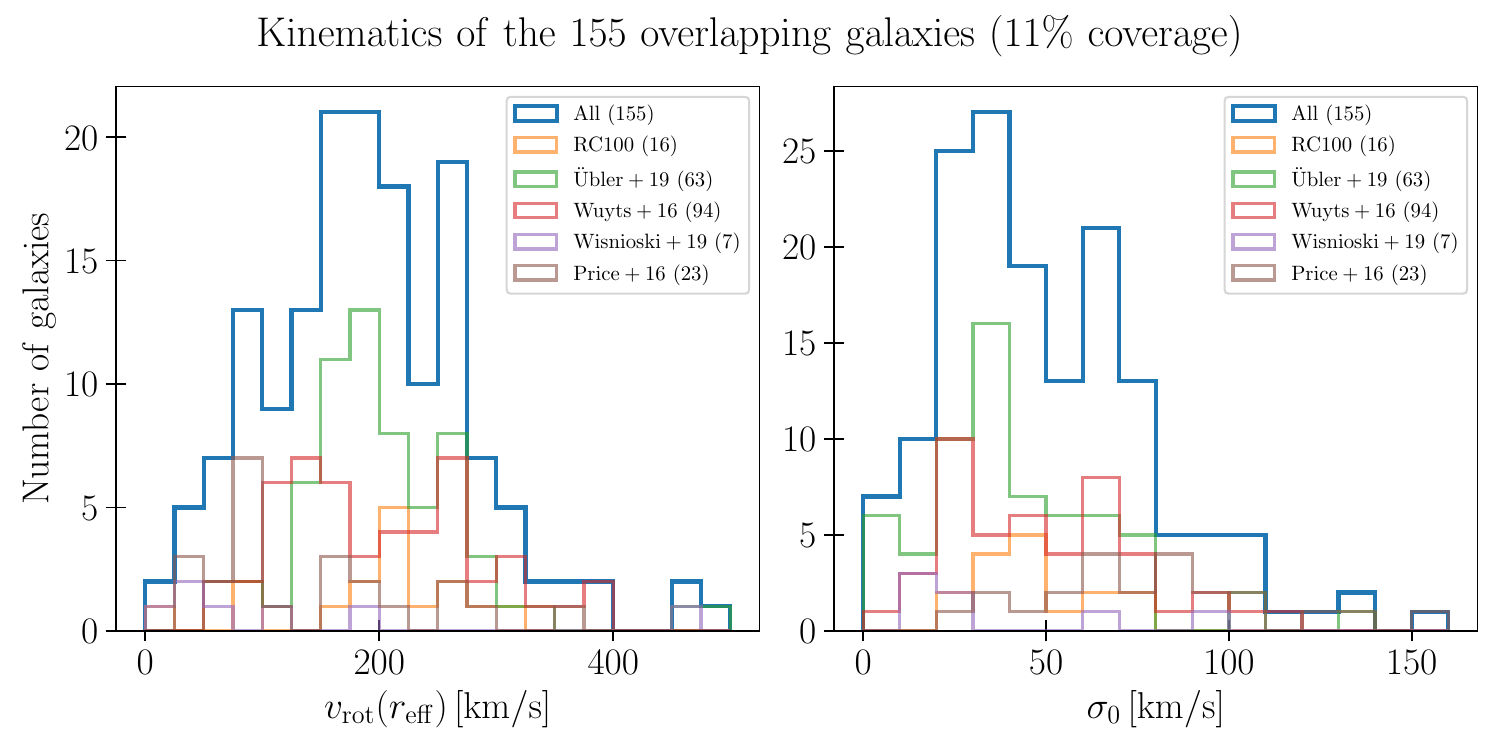}
	\caption{Histograms showing the distribution of rotational velocity $v_{\mathrm{rot}}(r_{\mathrm{eff}})$ (left) and velocity dispersion $\sigma_0$ (right) for different galaxy samples including the RC100 (\citealp{Nestor_2023}), and those from \cite{Uebler_2019}, \cite{Wuyts_2016}, \cite{Wisnioski_2019}, and \cite{Price_2016}. The legend indicates the corresponding dataset and includes the number of galaxies in each sample (in parentheses). The ``All'' sample (blue) represents the full dataset; the other colors correspond to specific subsamples from various studies.} 
    \label{fig: Overlap with kinematics}
\end{figure*}

\subsection{Reduction of imaging data}
\label{subsection: Data reduction}

We evaluated two distinct data reduction approaches: the public releases from the DAWN \JWST Archive\footnote{\url{https://dawn-cph.github.io/dja/}} reduction (v7.0 + v7.1) and JADES\footnote{\url{https://archive.stsci.edu/hlsp/jades}} (\citealp{Eisenstein_2023}) (v2.0) as well as our custom reduction pipeline. The DAWN Archive provides high-quality reductions of \JWST NIRCam data using the \texttt{grizli} reduction framework (\citealp{grizli}), which includes baseline calibrations, astrometric alignment using \texttt{DrizzlePac}, and background subtraction optimized for faint structure preservation and noise minimization. The JADES reduction products, while also high-quality, place greater emphasis on precise background modeling and outlier rejection via median mosaics, whereas \texttt{grizli} (DAWN) applies global sky corrections and sigma-clipping during stacking. For astrometric alignment, JADES employs \textsc{Montage3}, which offers precise registration across multiple exposures, differing from the \texttt{DrizzlePac}-based approach used in DAWN.

In parallel, we developed a custom reduction pipeline tailored to the specific requirements of our analysis. The reduction follows the standard \JWST image processing stages\footnote{\url{https://jwst-docs.stsci.edu/jwst-science-calibration-pipeline/stages-of-jwst-data-processing}}: \textit{i)} converting detector volt signals into rates of units of data numbers per second (DN/s); \textit{ii)} calibrating rates into flux densities and assigning world coordinates; and \textit{iii)} creating mosaic images from individual exposures while median-clipping outlier pixels. Building on previous studies (e.g., \citealt{Bagley_2023}; M. Franco et al. in prep.), we also: \textit{i)} improved removal of NIRCam imaging $1/f$ noise by detecting and subtracting horizontal and vertical stripes; \textit{ii)} enhanced removal of snowballs and claws through a combination of manual masking and template fitting, using the NIRCam team’s templates\footnote{\url{https://jwst-docs.stsci.edu/depreciated-jdox-articles/nircam-claws-and-wisps}. The latest templates were released recently (\protect\url{https://www.stsci.edu/contents/news/jwst/2024/new-nircam-wisp-templates-are-now-available}), but as our visually inspected targets are unaffected by claws, we did not update our reduction with the latest templates.}; \textit{iii)} improved world coordinate alignment (astrometry correction) by first detecting bright unsaturated stars and/or galaxies using \textsc{sextractor} (\citealp{Bertin_Arnout_96_sextractor}) and then cross-matching them with 3D-HST reference galaxy catalogs, ensuring alignment of \JWST images with 3D-HST images; \textit{iv)} additional flat background subtraction at the end of stage two using the Python \textsc{photutils} package before mosaicking to homogenize the mosaic background.

\begin{figure*}
\centering
	\includegraphics[width=0.95\textwidth]{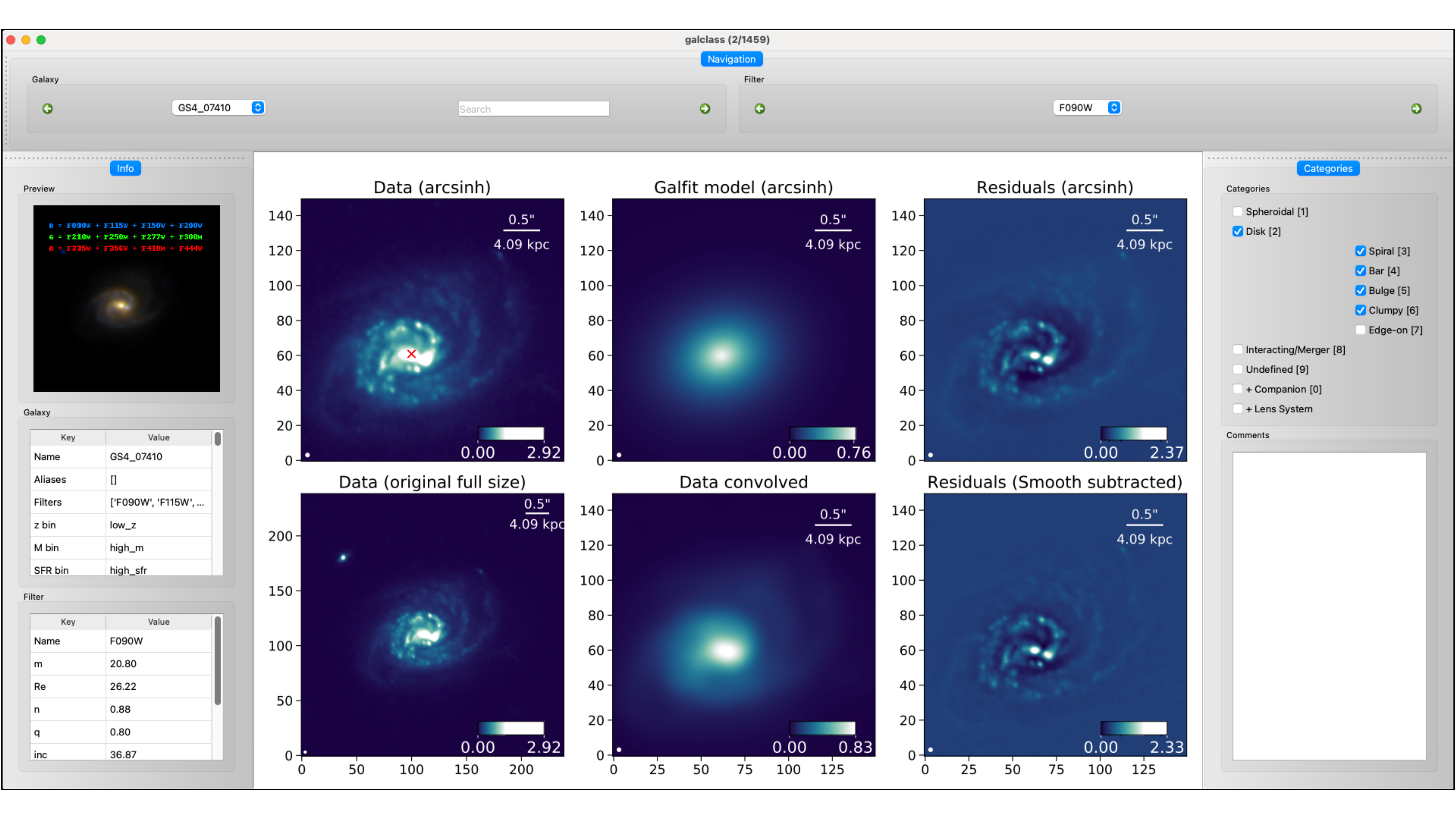}
	\caption{Example of the interactive tool \texttt{GalClass} (\citealp{Pastras_2025_galclass}) used for our morphological classification. The displayed galaxy, GS4\_07410, is shown in the F090W filter. The top row displays the original image, \galfit model, and residuals; the bottom row shows the full-size image, the smoothed image, and corresponding residuals. The left sidebar provides metadata and \galfit parameters, while the right panel contains the classification interface.}
    \label{fig: galclass example}
\end{figure*}

\begin{figure*}
\centering
	\includegraphics[width=0.99\textwidth]{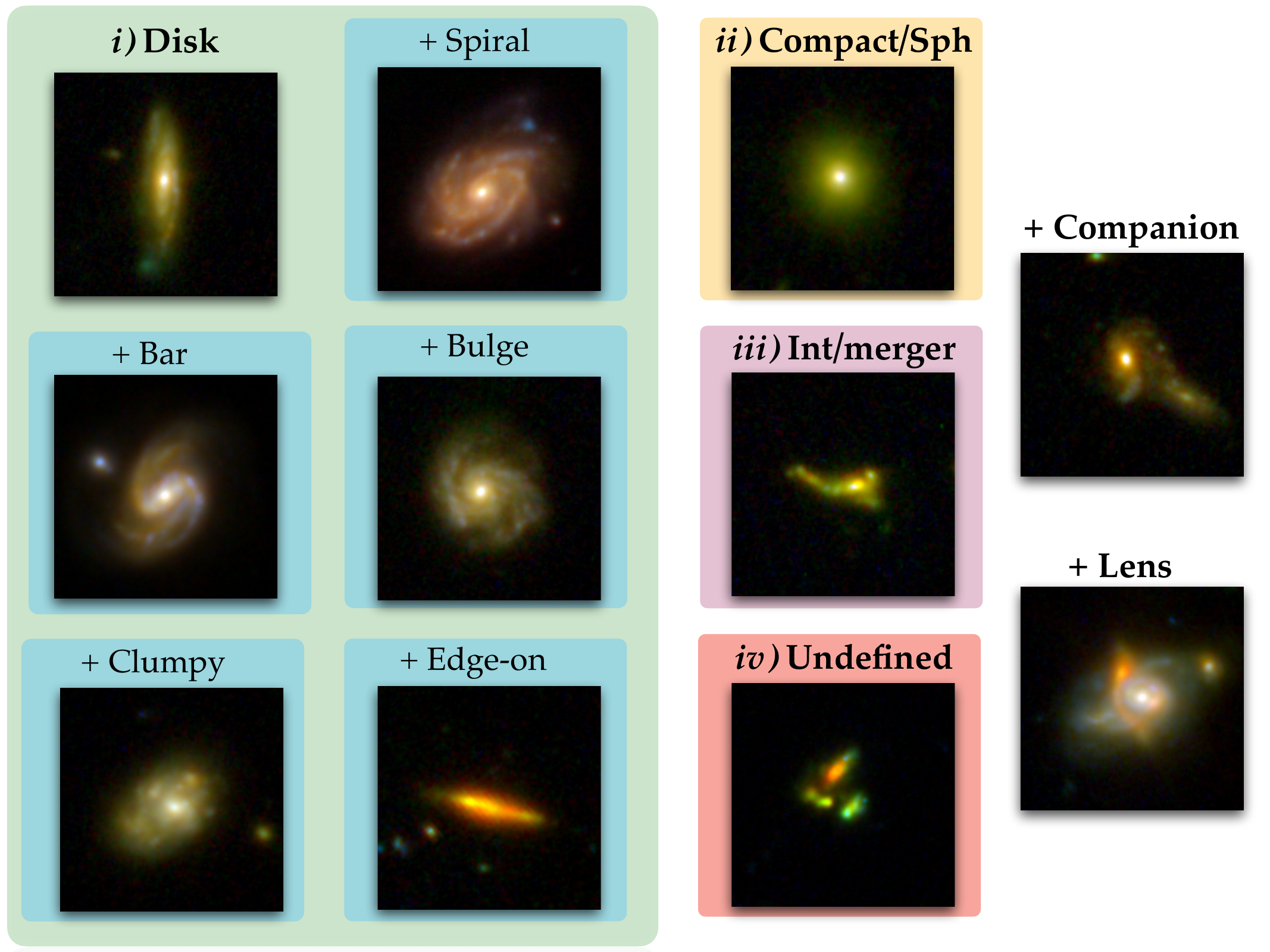}
	\caption{Classification scheme used in the visual classification, including examples from the galaxy sample analyzed in this study. The colored boxes, green, yellow, purple, and red, indicate the four main categories of disk, compact/spheroidal, interacting/merger, and undefined. The + Companion and + Lens classifications are nonmutually exclusive, meaning a galaxy could belong to any of the main categories, while also being identified as having a companion or a lensed system. The blue squares inside the disk category correspond to the five non-exclusive morphological features that a disk galaxy can have.}
    \label{fig: Classification scheme}
\end{figure*}

All scripts and manual mask regions are publicly available on GitHub\footnote{\url{https://github.com/1054/Crab.Toolkit.JWST}}. We processed the imaging data as soon as they became available in the \JWST MAST archive from various programs, frequently updating the Calibration Reference Data System (CRDS) and \JWST pipeline versions. For this study, the final products used for the morphology analysis were processed using CRDS calibration versions 1023--1130 and the \JWST pipeline versions 1.9.0--1.11.4. The final images were produced using a pixel scale of 25 mas, aligned with North up, to facilitate the morphological analysis. 

Prioritizing minimal noise, optimal background subtraction, and consistent PSF characteristics, we primarily adopted the DAWN reductions, which accounted for the majority of our sample ($\approx 96\%$). The JADES reductions were used for a small fraction of galaxies ($\approx 2\%$), while our custom reductions were employed in a few cases ($\approx 2\%$). Importantly, all three reduction pipelines produced high-quality data, ensuring that our final sample remained uniform and unaffected by these choices. To maintain consistency across filters, we ensured that all images for a given galaxy originated from the same reduction pipeline.

\begin{figure*}
\centering
	\includegraphics[width=0.95\textwidth]{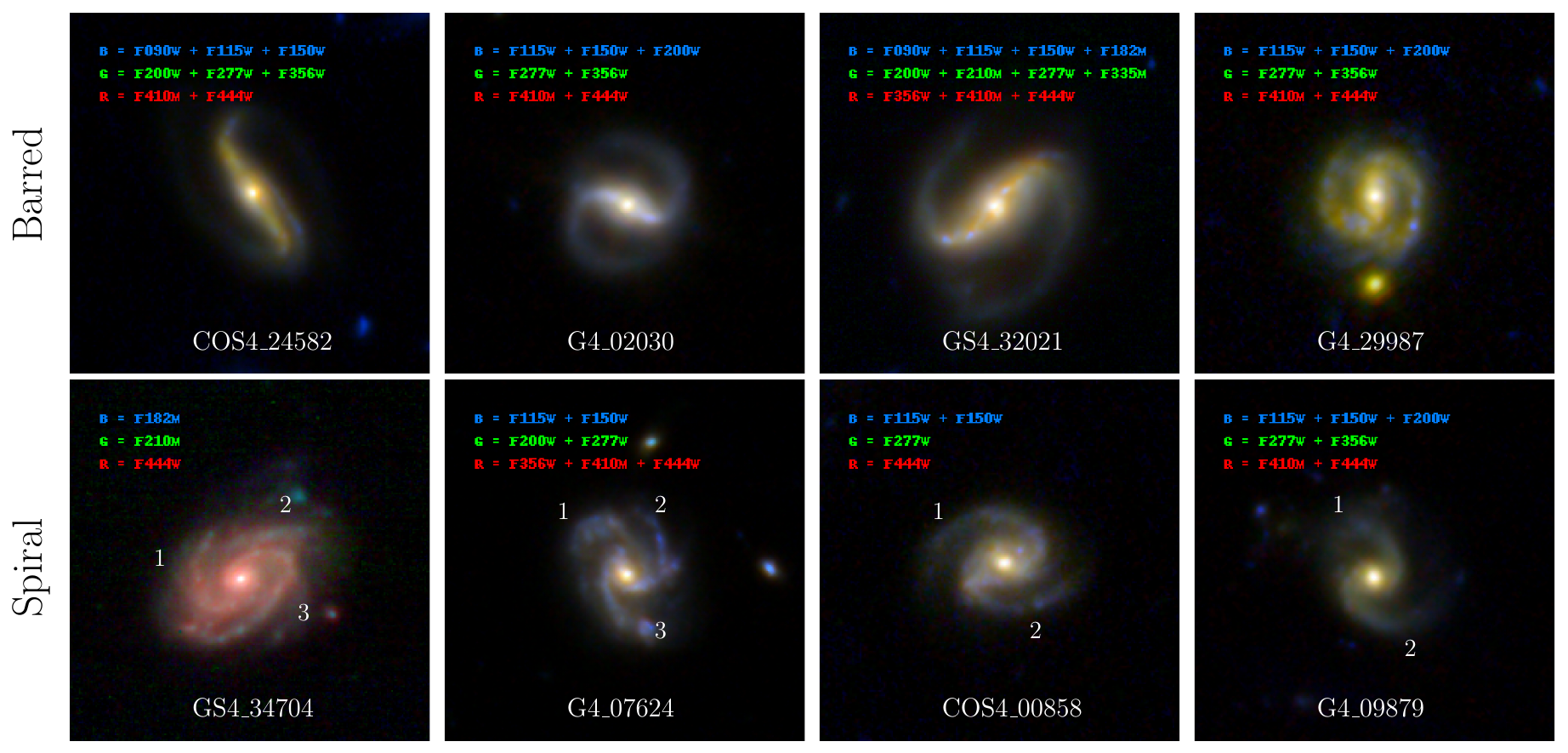}
	\caption{Example color composite images of galaxies classified as barred (top row) and spiral (bottom row).  These categories are nonmutually exclusive, as many barred galaxies also exhibit clear spiral arms. The filters used to create each composite image are labeled in the top left corner of each panel. The numbers in the bottom row indicate the number of spiral arms identified in the visual classification. The full mosaics of the barred and spiral galaxies can be found at the following two links: \href{https://www.mpe.mpg.de/resources/IR/JWSTColorImages/barred_grid_A4.pdf}{bars} and \href{https://www.mpe.mpg.de/resources/IR/JWSTColorImages/spirals_grid_A4_merged.pdf}{spirals}.}
    \label{fig: Examples of bars and spirals}
\end{figure*}

\subsection{Kinematics sample}
\label{subsection: Kinematics sample}

To trace the dynamical evolution of galaxies at the epochs probed in this study, we also examined the interplay between morphological features observed in the \JWST NIRCam images and available kinematic measurements, specifically the $v/\sigma$ ratio, which provides crucial insights into each galaxy’s dynamical state. To compile the kinematic data, we used measurements from well-established surveys and studies. 

The KMOS$^{\rm 3D}$ survey (\citealp{KMOS3D}) is the largest and deepest IFU survey in the deep fields, covering a redshift range of $z\sim 0.6-2.6$. This survey has provided key insights into the kinematic properties of star-forming galaxies across cosmic time, utilizing high-resolution H$\alpha$ emission line maps obtained with KMOS (K-band Multi-Object Spectrograph) on the VLT. Multiple studies have explored the kinematics of these galaxies using a variety of techniques, including disk modeling, kinematic classification, and dynamical mass estimates \citep[e.g.,][]{Burkert, Wuyts_2016, Genzel_2017, Ubler_2017, Lang_2017, Ubler_2018, Wisnioski_2019, Forster_2019}. These analyses have provided critical constraints on the evolution of angular momentum, velocity dispersion, and the transition between rotation- and dispersion-dominated systems. Beyond KMOS$^{\rm 3D}$, we incorporated high-resolution kinematic measurements for 100 galaxies (RC100) based on H$\alpha$ and CO tracers from \mbox{\citet{Nestor_2023}}, as well as slit-spectroscopy kinematics from \citet{Price_2016}. For galaxies that have multiple kinematic measurements, we prioritized data sources based on methodological advancements and data quality, following this order: RC100 \citep{Nestor_2023}, followed by KMOS$^{\rm 3D}$-based studies \citep{Uebler_2019, Wuyts_2016, Wisnioski_2019}, and finally, the slit-spectroscopic MOSDEF survey \citep{Price_2016}.

We found an overlap of 155 galaxies that have velocity and velocity dispersion measurements. For consistency, we adopted a standardized approach, using the rotation velocity at one effective radius, $v_\mathrm{rot}(\re)$, and assume a radially constant intrinsic disk velocity dispersion $\sigma_0$, corrected for observational effects, such as beam smearing, as derived from modeling. The detailed measurement adjustments are outlined as follows:

\begin{itemize}
    \item For RC100 \citep{Nestor_2023}, where we included 16 out of their 100 galaxies, we used reported circular velocities and applied the following correction to obtain $v_\mathrm{rot}(\re)$: 
    \begin{equation}
    v_\mathrm{rot}^2(r) = v_\mathrm{circ}^2(r) - 3.36 \left(\frac{r}{\re}\right) \sigma_0^2
    \end{equation}

    \item In \cite{Uebler_2019}, we identified an overlap of 71 galaxies from their sample of 187; however, based on our prioritization, and since some of these are already in the RC100 sample, we included 63. The study reported values for $v_\mathrm{rot}(r=1.38 \re)$, which we adjusted to $v_\mathrm{rot}(\re)$ using rotation curve models for a thick disk as described by \cite{Noordermeer_2008}.
 
    \item For \cite{Wuyts_2016}, out of a sample of 240, we used 50 galaxies based on prioritization criteria. 
    
    \item In \cite{Wisnioski_2019}, from a total number of 745, we identified seven more overlapping galaxies.

    \item Lastly, for \cite{Price_2020} (MOSDEF), we identified 26 overlapping galaxies from their full sample of 1,717, of which we used 23. Since these measurements were based on slit spectroscopy from the MOSFIRE instrument, the slits could not be precisely aligned with the galaxies' major axes. The sample of 23 used here is restricted to objects that are spatially resolved (under seeing-limited conditions), have reasonable alignment between slit and galaxy major axes ($|\Delta\mathrm{PA}|\leq45^{\circ}$), and have detected rotation signatures (within the uncertainties). We note that this work used an arctan parametric fit to derive the rotation velocity profile, differing from the parametric fits used in other surveys. Moreover, as the exact kinematic major axes were unknown, the modeling assumed the morphological and kinematic axes were the same, introducing potential systematics. However, given the small sample size, any such effect is unlikely to have a significant impact on our overall analysis.

\end{itemize}
    
An overview of the distributions of velocities and velocity dispersions is shown in Figure \ref{fig: Overlap with kinematics}.

\begin{figure*}
\sidecaption
  \includegraphics[width=12.9cm]{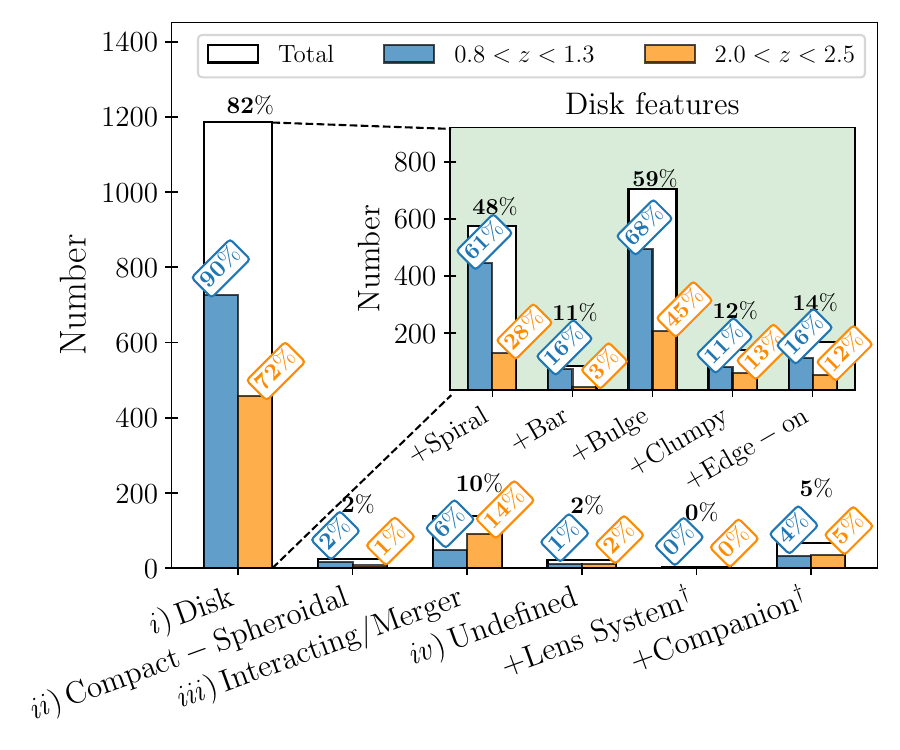}
     \caption{Global classification results for the 1,451 galaxies in our sample. The blue and orange bars represent the low- and high-redshift bins, respectively, while the white bars correspond to the full sample. The percentages on top of the bars correspond to fractions normalized by the number of galaxies in the full sample, where black indicates the total, and blue and orange the two redshift bins. While the main text reports fractions with associated uncertainties, we show percentages in this figure to facilitate readability and visual comparison across bins. The inset figure highlights the galaxies classified as disks in the sample, for which we also note the presence of five morphological features. The percentages displayed on top of those bars indicate the fractions of each subcategory normalized by the total number of disks. The $\dagger$ symbol in + Companion denotes a nonmutually exclusive category, meaning it can be assigned alongside any of the main categories.}
     \label{fig: Classification results}
\end{figure*}

\begin{table*}
    \caption{Classification results overview}
    \label{tab: Main classifications}
    \centering
\begin{tabular}{ccccccccccc}
\hline\hline
\multirow{2}{*}{Type}      &  \multirow{2}{*}{$f$ definition}  & \multicolumn{2}{c}{Total} &  & \multicolumn{2}{c}{$0.8<z<1.3$} & & \multicolumn{2}{c}{$2.0<z<2.5$} \\
\cline{3-4} \cline{6-7} \cline{9-10}
              &                                            & $N$     &$f$(\%)    & & $N$   &$f$(\%)    & & $N$  &$f$(\%)   \\
\hline
Total         &  $N_{\rm total}/N_{\rm total}$             & 1,451   & 100       & & 811   & 100       & & 640  & 100      \\
Disk-like     &  $N_{\rm disk}/N_{\rm total}$              & 1,186   & $82\pm3$   & & 727   & $90\pm3$   & & 459  & $72\pm5$  \\
Compact/Sph   &  $N_{\rm c,s}/N_{\rm total}$               & 24      & $2\pm1$    & & 17    & $2\pm2$    & & 7    & $1\pm1$   \\
Int/Merger    &  $N_{\rm i,m}/N_{\rm total}$               & 139     & $10\pm2$   & & 48    & $6\pm2$    & & 91   & $14\pm4$  \\
Undefined     &  $N_{\rm undefined}/N_{\rm total}$         & 22      & $2\pm1$    & & 11    & $1\pm1$    & & 11   & $2\pm2$   \\
+ Lens systems    & $N_{\rm lens}/N_{\rm total}$           & 3       & $<1\pm0$   & & 1     & $1\pm0$    & & 2    & $<1\pm1$  \\
+ Companion    & $N_{\rm + \, companion}/N_{\rm total}$    & 67      & $5\pm2$    & & 32    & $4\pm2$    & & 35   & $6\pm3$   \\
No 4/6 agreement &  $N_{\rm no-agree}/N_{\rm total}$       & 80      & $6\pm2$    & & 8     & $1\pm1$    & & 72   & $11\pm4$  \\
Spirals        &  $N_{\rm spiral}/N_{\rm total}$           & 574     & $40\pm4$   & & 444   & $55\pm5$   & & 130  & $20\pm5$  \\
Bars   &  $N_{\rm bar}/N_{\rm total}$                      & 43      & $3\pm1$    & & 42    & $5\pm2$    & & 1    & $<1\pm0$  \\
Bulges         &  $N_{\rm bulge}/N_{\rm total}$            & 703     & $48\pm4$   & & 495   & $61\pm5$   & & 208  & $33\pm6$  \\

\hline

Spirals        &  $N_{\rm spiral}/N_{\rm disk}$            & 574      & $48\pm4$  & & 444   & $61\pm5$   & & 130  & $28\pm6$  \\
Bars &  $N_{\rm bar}/N_{\rm disk}$                         & 43       & $4\pm2$   & & 42    & $6\pm3$    & & 1    & $<1\pm1$  \\
Bars (isophotal fitting) &  $N_{\rm bar}^{(a)}/N_{\rm disk}^{(b)}$ & 84 & $11\pm3$& & 74    & $16\pm5$   & & 10   & $3\pm3$   \\
Bulges         &  $N_{\rm bulge}/N_{\rm disk}$             & 703      & $59\pm4$  & & 495   & $68\pm5$   & & 208  & $45\pm7$  \\
Clumpy         &  $N_{\rm clumpy}/N_{\rm disk}$            & 140      & $12\pm3$  & & 81    & $11\pm4$   & & 59   & $13\pm5$  \\
Edge-on        &  $N_{\rm edge-on}/N_{\rm disk}$           & 167      & $14\pm3$  & & 113   & $16\pm4$   & & 54   & $12\pm5$  \\

\bottomrule

Two-arm spirals   &  $N_{\rm 2,arm}/N_{\rm spiral}^{(c)}$  &  208   & $61\pm8$    & &  156  &  $60\pm9$  & &  52  & $64\pm16$  \\
Three-arm spirals &  $N_{\rm 3,arm}/N_{\rm spiral}^{(c)}$  &  104   & $30\pm7$    & &  79   &  $30\pm9$  & &  25  & $31\pm15$  \\
Four-arm spirals  &  $N_{\rm 4,arm}/N_{\rm spiral}^{(c)}$  &  18    & $5\pm4$     & &  15   &  $6\pm4$   & &  8   & $4\pm6$   \\
\end{tabular}
\tablefoot{
         All the fractions above are based on our 4/6 expert vote threshold. We show the main morphological categories with their distribution (in percentage) across redshift bins and corresponding $3\sigma$ standard errors. The top rows list total fractions normalized by the full sample, while the middle section details subcategories among disk galaxies. The bottom rows present the distribution of spiral arm counts.\\
        \tablefoottext{a}{These values correspond to the classification using isophotal fitting on a larger candidate sample, selected with a 2/6 voting threshold and further assessed as detailed in Sect. \ref{subsection: Classification overview}.}\\
        \tablefoottext{b}{These values correspond to those with an ellipticity threshold of $e<0.5$.}\\
       \tablefoottext{c}{$N_\mathrm{spiral}$ in the bottom rows corresponds to the number of galaxies for which a 4/6 agreement was reached (341 galaxies).}}
\end{table*}

\section{Visual classifications of morphology}
\label{section: Galaxy morphology}

\subsection{Classification scheme}
\label{subsection: Classification scheme}

To prepare the \JWST NIRCam imaging for analysis, we first fitted the images using a single \sersic profile with \galfit (\citealp{Peng_2002, Peng_2010}) for each stamp. Initial source detection relied on segmentation maps, which provided preliminary inputs for the \galfit routine. Following this, six of the authors of this work (J.M.E.S, C.B, S.P, C.P, J.C, J-B.J) visually classified all the galaxies in the sample. For increased robustness, the morphological classification process incorporated available image stamps, color composites, residual images, and the best-fit \sersic indices. Additionally, unsharp-masked versions of the images were also shown to the classifiers to enhance fine details and highlight distinct features that might otherwise be overlooked. To facilitate the classification process, given the large sample size of approximately 10,000 images, we used a custom classification tool for galaxy morphologies (\texttt{GalClass}; \citealp{Pastras_2025_galclass})\footnote{Available at \url{https://github.com/spastras/galclass}}.

Our visual classification categorizes the galaxies in the sample based on their morphological features, emphasizing structural characteristics such as compactness, diskiness, interaction signatures, and lensing distortion. Below, we outline the main categories used in the classification:\\

\textit{i)} Disk: These galaxies are characterized by a distinct, organized structure with a S\'ersic index close to unity, consistent with \cite{Wuyts_2011}, who showed that $n \sim 1$ is typical for main-sequence galaxies. They often show spiral arms and may have a central bulge. Depending on the viewing angle, they can appear circular, oval, or linear, corresponding to face-on, intermediate, or edge-on orientations.\\
\textit{ii)} Compact-Spheroidal: These systems appear smooth and rounded, and show no visible disk or spiral structure. They have a uniform brightness profile and are typically featureless. When classifying these galaxies, we also considered their \sersic indices, selecting systems that were well described by $n \approx 4$.\\
\textit{iii)} Interacting/merger: Interacting or merging galaxies exhibit irregular and distorted shapes due to gravitational interactions. Common features include tidal tails, bridges between galaxies, and unusual shapes, often indicating ongoing collisions or close proximity to another galaxy.\\
\textit{iv)} Undefined: This category includes galaxies that lack any well-defined structure but do not show clear interaction features. They appear disordered, fragmented, or amorphous, without the characteristics of disks, spheroids, or mergers.\\

Within each category, we also included an option to identify the presence of companion galaxies as well as a strongly lensed system, identified by the presence of arcs, or rings of a background object gravitationally lensed by the foreground galaxy. For galaxies classified as disks, we also noted the presence or not of five non-exclusive morphological features labeled Spiral, Bar, Bulge, Clumpy, and Edge-on, several of which may be present in a given galaxy. Figure \ref{fig: galclass example} shows an example of \texttt{GalClass} for a spiral galaxy, and Figure \ref{fig: Classification scheme} provides an overview of the classification scheme, showcasing the main categories and representative examples of each category and subcategory with color composite images. These images were generated using \textsc{trilogy} \citep{Coe_2012}, using filter groupings carefully selected to enhance the visibility of key galaxy features. Specifically, shorter-wavelength filters were assigned to blue, longer-wavelength filters to red, and intermediate-wavelength filters to green, ensuring a balanced color representation. An equal number of filters (where possible) were distributed across the RGB channels to maintain consistency in the visualization. \footnote{The individual images used to generate the color composites were not PSF-matched. While variations in PSF size exist across filters, the impact on the color images is generally minor, as the broad structural features of galaxies remain visually consistent. Moreover, these images are intended solely for visual representation of the sample, and no quantitative measurements are derived from them.}

The final classification was determined based on a consensus threshold, requiring at least four out of six expert votes for a category assignment. We emphasize that galaxies failing to meet this threshold for any category remain unclassified, which is distinct from the ``Undefined'' category, as the latter represents galaxies that have inherently disordered shapes or noncoherent structures, and is thus a category on its own. Figure \ref{fig: Examples of bars and spirals} 
illustrates representative cases of galaxies identified as bars and spirals, each meeting the consensus threshold of four out of six expert votes.

From this point onward, all quoted fractions in the text and plots will be accompanied by their corresponding uncertainties, computed as 3$\sigma$ standard errors derived from binomial statistics (unless otherwise noted). Specifically, we adopted $\sigma = \sqrt{p(1 - p)/N}$, where $p$ is the measured fraction and $N$ is the total number of galaxies in the corresponding parent sample. For small subsamples, this approach may slightly overestimate the uncertainty; however, we applied it uniformly for consistency and comparability across all measurements.

\subsection{Classification overview}
\label{subsection: Classification overview}

Our visual classification results, based on the 4/6 voting threshold (Table \ref{tab: Main classifications} and Figure \ref{fig: Classification results}), reveal that the majority of galaxies in our sample have a disk-like morphology, comprising a fraction of $0.82\pm 0.03$ of the total sample ($0.90\pm 0.03$ and $0.72\pm 0.05$ in the low- and high-redshift bins, respectively). Interacting or merging galaxies form the second-largest category, accounting for approximately $0.10\pm 0.02$ of the sample, and showing a notable increase in the higher-redshift bin. Spheroidal galaxies make up a small fraction ($0.02\pm 0.01$), with slightly more found in the lower-redshift bin. Lensed systems and undefined galaxies are rare, each contributing less than 0.01. Notably, only 80 galaxies (fraction $\approx 0.06\pm 0.02$) lacked a 4/6 voting agreement.

\begin{figure}
    \centering
    \includegraphics[width=0.5\textwidth]{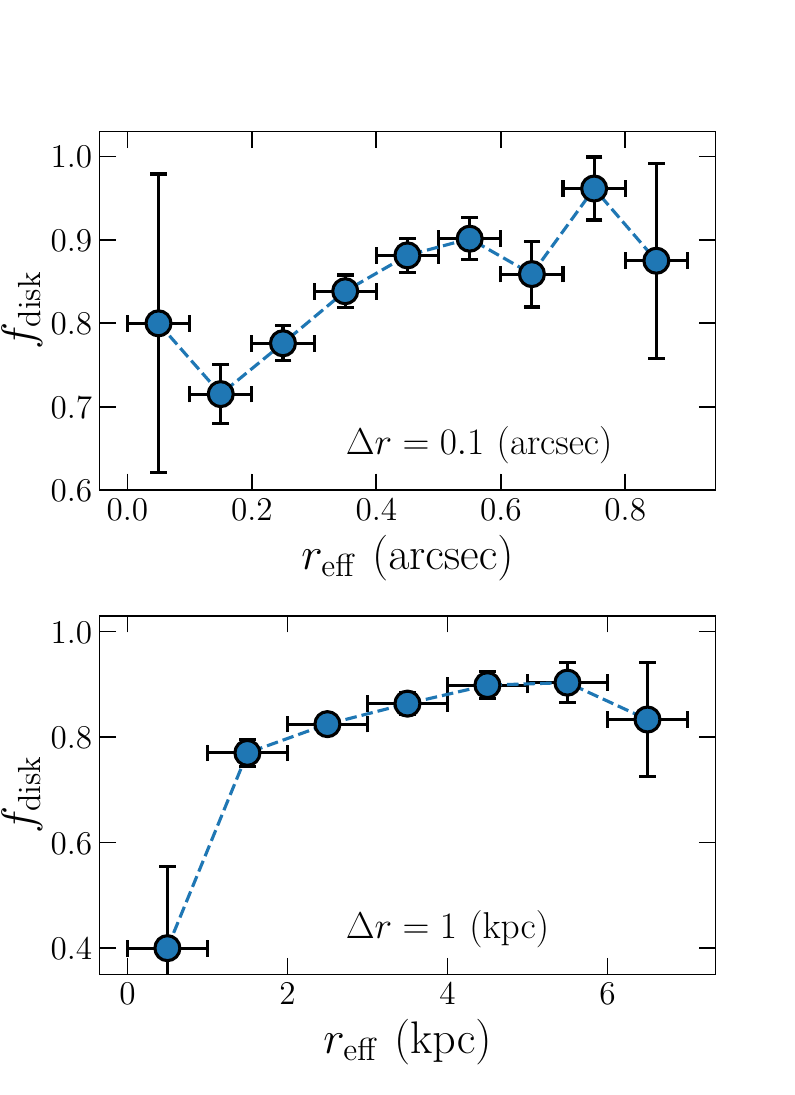}
    \caption{Disk fraction as a function of effective radius $r_\mathrm{eff}$, measured in angular units with bin size $\Delta r = 0.1$ arcsec (top) and in physical units of $\Delta r = 1$ kpc (bottom). Blue points represent the disk fraction in each bin, and error bars indicate binomial uncertainties. Dashed lines are shown to guide the eye. In both panels, the disk fraction increases with effective radius, highlighting that disks are more easily identified in galaxies with larger apparent and physical sizes.}
    \label{fig: Disk fraction vs effective radius}
\end{figure}

\begin{figure*}
\centering
\begin{subfigure}{0.95\textwidth}
    \centering
    \includegraphics[width=\textwidth]{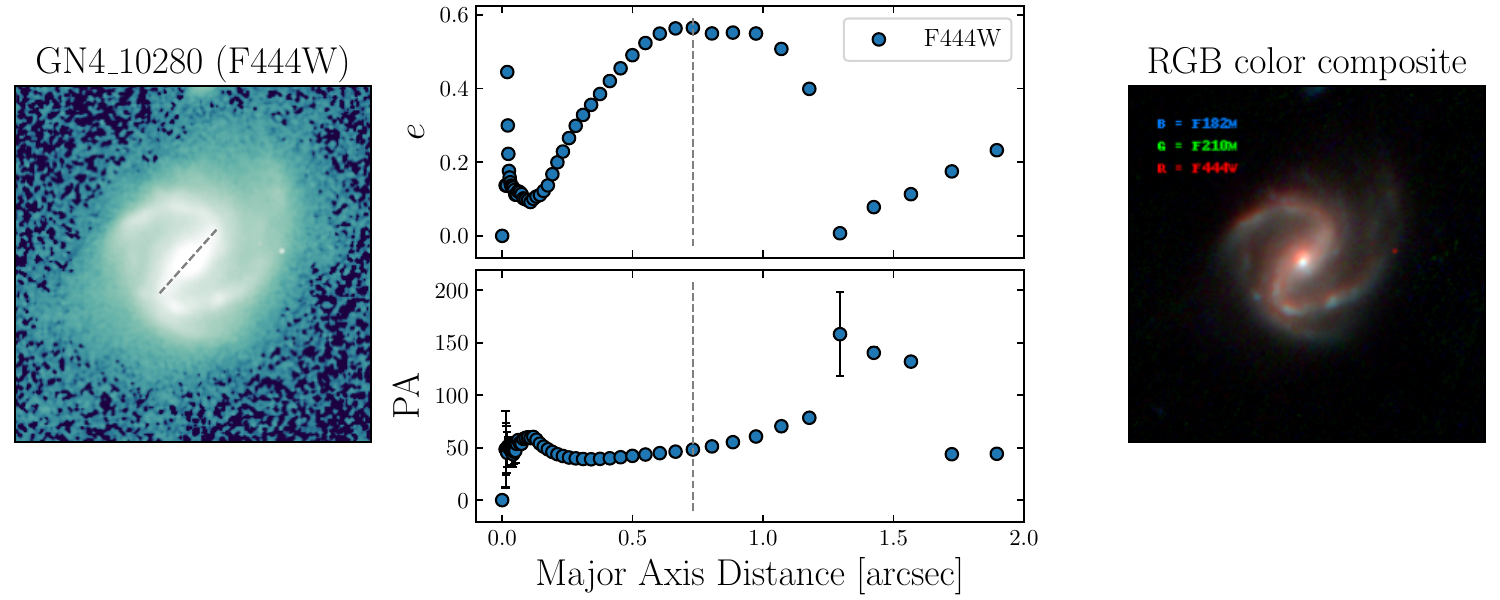}
    \label{fig:bar_identification_top}
\end{subfigure}

\vspace{0.5em} 

\begin{subfigure}{0.95\textwidth}
    \centering
    \includegraphics[width=\textwidth]{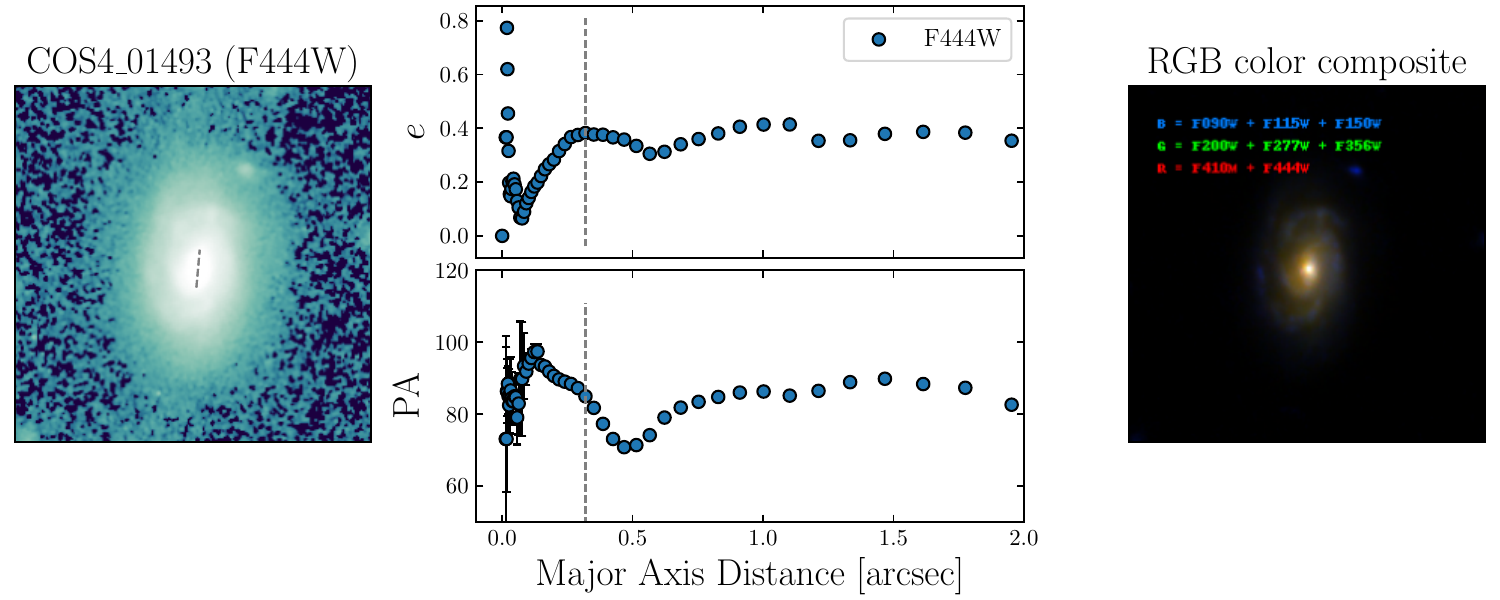}
    \label{fig:bar_identification_bottom}
\end{subfigure}

\caption{
Bar identification analysis for two example galaxies. 
\textit{Top:} GN4\_10280, identified as barred based on both visual morphology (N=6 votes) and quantitative isophotal fitting. Left panel: F444W image with best-fit bar major axis (dashed). Middle: radial profiles of ellipticity ($e$) and position angle (PA) with the ellipticity peak marked. Right: RGB composite showing structural features.  
\textit{Bottom:} COS4\_01493, initially not classified as barred via visual inspection (N=3 votes), but later confirmed as barred through isophotal fitting.
}
\label{fig: bar identification}
\end{figure*}

An important potential bias to consider in the classification of galaxies (whether through visual or automatic methods) is related to their apparent size. Specifically, we observe increasing disk fractions $f_\mathrm{disk}$ as a function of size as shown in Figure \ref{fig: Disk fraction vs effective radius}, where we show $f_\mathrm{disk}$ as a function of effective radius in apparent size (arcsec) and physical size (kpc). This is expected since it is easier to identify a disk-like structure in extended sources. However, this trend may also have a physical origin, as disks typically exhibit extended light distributions, which naturally result in larger effective radii and contribute to the observed correlation. An additional discussion on observational limitations, such as spatial resolution, surface brightness dimming, and wavelength coverage, is provided in Appendix \ref{appendix - Observational limitations}.

\subsection{Classification of substructures}

\subsubsection{Spiral structure}
\label{subsection: Spiral structure}

Despite the lack of consensus on the origin of spiral structure, its connection to the dynamical state of the host galaxy is well supported (see reviews by \citealp{Dobbs_2014, Shu_2016, Sellwood_2022}). Spiral structure arises in disks that are sufficiently cold and rotationally supported, offering an indirect tracer of a galaxy’s mass distribution and instability to non-axisymmetric perturbations (e.g., \citealp{Athanassoula_1984, Toomre, Sellwood_1984}). Identifying spiral arms, therefore, indicates that the galaxy hosts a settled disk, responsive to internal or external perturbations (e.g., \citealp{Dobbs_2014}). 

Beyond their presence alone, the number of spiral arms also encodes information about the underlying dynamics (e.g., \citealp{Elmegreen_2014, Dobbs_2014}). While two-armed (grand-design) spirals are likely driven by global disk modes or external perturbations, such as bars or tidal interactions (e.g., \citealp{Toomre_1972, Athanassoula_1984, Dobbs_2010, D'Onghia_2013, Kendall_2015}), multi-armed (flocculent) spirals typically emerge in marginally stable disks through mechanisms such as swing amplification (\citealp{Toomre_1981}). As such, both the detection and characterization of spiral arms (particularly arm multiplicity) offer key diagnostics of disk structure, stability, and the mechanisms shaping galaxy evolution (e.g., \citealp{Hart_2017}).
    
To investigate potential trends in spiral structure across our two redshift bins, we visually counted the number of arms within the spiral galaxies identified in our classification (574 in total). Following a similar approach used for morphological classification, six authors (J.M.E.S., C.B., C.P., J.C., J-B.J., G.T.) independently performed the arm counts. To aid in the identification of spiral arms, we utilized a custom interactive tool for visualizing FITS files with customizable scales using all the stamps for each galaxy. Each team member independently determined the number of visible arms for each galaxy. Figure \ref{fig: Examples of bars and spirals} (bottom panels) provides examples of galaxies with consensus arm counts, while full-color composite images of all classified spiral galaxies are available at \href{https://www.mpe.mpg.de/resources/IR/JWSTColorImages/spirals_grid_A4_merged.pdf}{this link}. The results of the spiral arm counting are discussed in Sect. \ref{section: Morphological evolution}.

\subsubsection{Bar structure}
\label{subsection: Bars}

We identified barred galaxies in our sample by refining the initial visual classification through the fitting of elliptical isophotes. Visually identifying bars can be challenging as they can be obscured by bright central bulges or may blend with spiral arms. Therefore, we applied a more inclusive threshold of 2/6 classifier votes to ensure potential barred galaxies were not overlooked. The isophotal fitting method provides a quantitative test for confirming bar structures.

Starting from a parent sample of 123 galaxies that received at least 2/6 classifier votes for a bar, we applied an ellipse-fitting technique using \texttt{photutils} (\citealp{Bradley_2016}) on images taken with the longest available wavelength filter (F444W, except for one case where F356W was used). We then analyzed the resulting ellipticity ($e = 1 - q$), where $q$ is the axis ratio, as well as the position angle profiles as functions of radius to confirm the presence of bars. To classify a galaxy as barred, we imposed the following criteria: \textit{i)} the galaxy must not be edge-on, defined as having an ellipticity below 0.5 at large radii, outside the regions influenced by the bar or spiral arms; \textit{ii)} The ellipticity must increase smoothly with radius, reaching a well-defined peak above 0.25 before decreasing again, as expected for a bar. This ensures that the feature is significantly elongated and not an artifact of noise or a mild bulge elongation; and \textit{iii)} the position angle must remain relatively constant (within 20°). We show two examples of this analysis in Figure \ref{fig: bar identification}: galaxy GN4\_10280, which received N=6 votes for barred system, and galaxy COS4\_01930, which only received N=3 votes but was later confirmed barred following this quantitative analysis.

Using these additional criteria, the number of identified barred systems increased to 84, where 74 are in the low-redshift bin and ten are in the high-redshift bin. Notably, 81 out of the 84 barred galaxies were also classified as spiral galaxies. Figure \ref{fig: Examples of bars and spirals} (top panels) presents four examples of barred galaxies in our sample. The full mosaic of the color images of the barred systems is available at \href{https://www.mpe.mpg.de/resources/IR/JWSTColorImages/barred_grid_A4.pdf}{this link}\footnote{\url{https://www.mpe.mpg.de/resources/IR/JWSTColorImages/barred_grid_A4.pdf}}.

\section{Morphological evolution}
\label{section: Morphological evolution}

In this section, we present and interpret the results of our visual classification of morphologies and a discussion on the various trends across redshift.

\begin{figure}
	\includegraphics[width=0.49\textwidth]{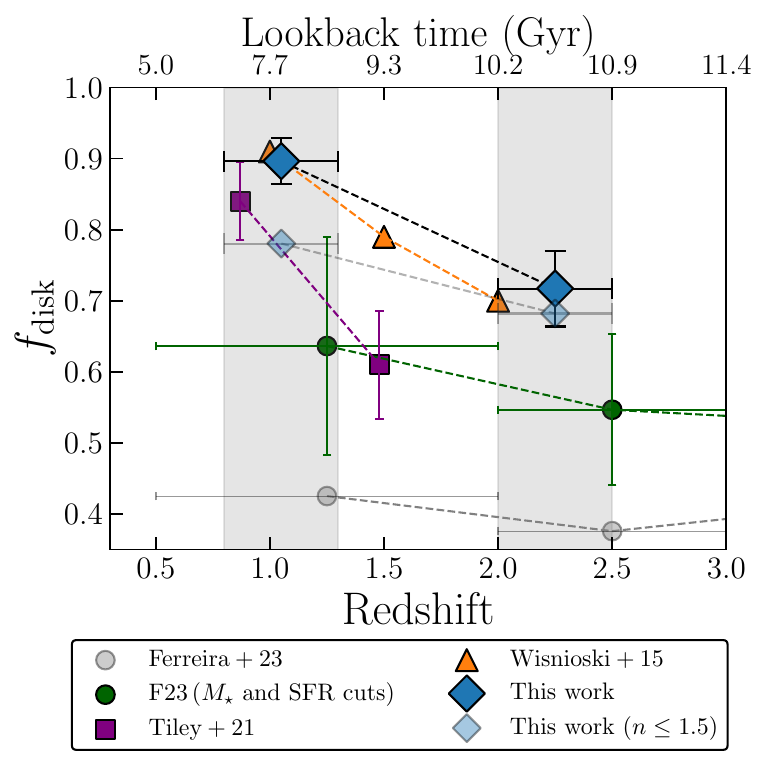}
	\caption{Disk fractions, $f_\mathrm{disk} = N_\mathrm{disk}/N_\mathrm{total}$, as a function of redshift. The dark blue diamonds represent the results from our visual classifications. The light blue diamonds indicate disk fractions when an additional S\'ersic index threshold of $n \leq 1.5$ is applied. For comparison, the orange triangles indicate disk fractions from the kinematic analysis of \protect\cite{KMOS3D}, while the purple squares show measurements from \protect\cite{Tiley_2021}. The green circles represent disk fractions for the subset of the 3D-HST galaxies in \protect\citep{Ferreira_2023} (F23) that overlap with our sample and satisfy our $M_{\star}$ \& SFR selection criteria. The gray circles correspond to the unfiltered but still overlapping sample of F23. The gray-shaded regions indicate the redshift bins analyzed in this study.}
    \label{fig: Disk fractions}
\end{figure}

\begin{figure}
    \centering
        \includegraphics[width=0.5\textwidth]{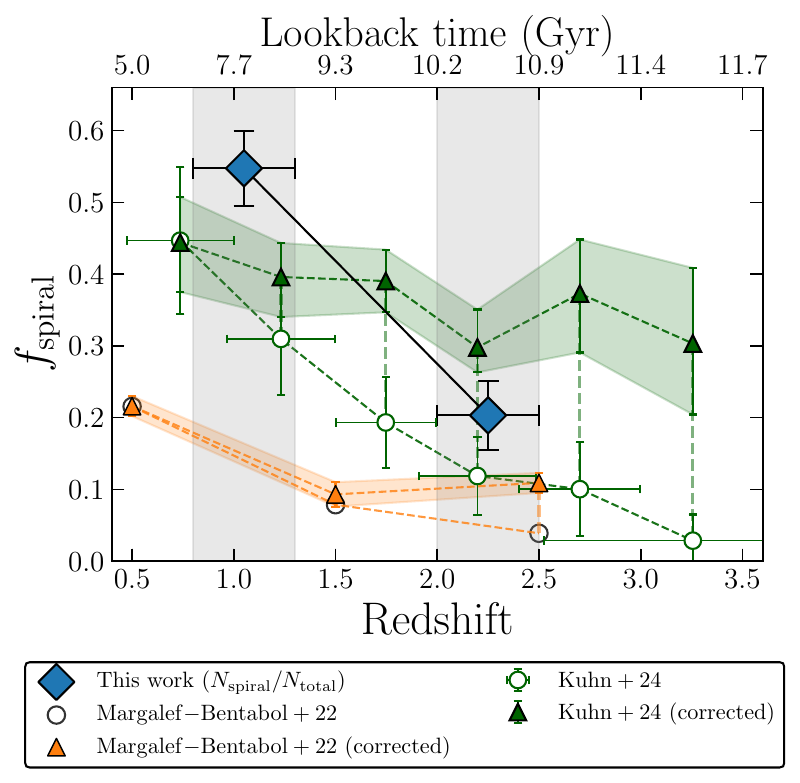}
        \caption{Spiral fractions defined as $f_\mathrm{spiral} = N_\mathrm{spiral}/N_\mathrm{total}$ for comparison with other studies. Green dots and triangles represent the \textit{JWST}-based results from \protect\citep{Kuhn_2024}, where the white circles show the directly measured spiral fractions, and the filled triangles indicate the values after applying corrections for observational effects. Orange dots represent the fractions inferred from \HST imaging in the CANDELS fields by \protect\citep{Margalef-Bentabol2022}.}
        \label{fig: Spiral fractions}
\end{figure}

\subsection{Disks}
\label{subsection: Disks evolution}

Figure \ref{fig: Disk fractions} compares the disk fractions $f_\mathrm{disk} = N_\mathrm{disk}/N_\mathrm{total}$ derived from our study, with the kinematic analysis of KMOS$^\mathrm{3D}$ \citep{KMOS3D} and a visual morphological classification by \protect\cite{Ferreira_2023}. We found that our large disk fractions ($0.90\pm0.03$ and $0.72\pm0.05$ in the low- and high-redshift bins, respectively) are consistent with the disk fractions estimated from the kinematic analysis of \citep{KMOS3D} across the same redshift range. Additionally, our low-redshift results are also consistent with the findings of \cite{Tiley_2021} (KGES) for their sample of 288 galaxies at $1.2<z<1.8$, where the $z\approx 0.9$ measurement comes from KROSS (\citealp{Harrison_2017}) with $v/\sigma>1$.

However, discrepancies arise when comparing our results to the visual classification results of \cite{Ferreira_2023}, even after applying similar stellar mass and star formation rate (SFR) cuts. These differences are likely driven by variations in classification schemes and sample selection criteria. In particular, Ferreira et al. studied a broader range of stellar masses, including low-mass systems within $\log(M_\star/M_\odot) \approx [6.6, 11.5]$ in the range $1.5<z<6.5$ for \JWST CEERS observations overlapping with the CANDELS EGS field. Our sample is much more massive with $\log(M_\star/M_\odot) = [10, 11.4]$. When applying comparable mass cuts, the sample overlap is very limited: we identified only 44 galaxies in the similar low-redshift bin and 75 in the high-redshift bin, where 28 and 41 are classified as disks, respectively.

Additionally, differences in classification criteria may contribute to the observed discrepancies. A key distinction in their approach is the ``peculiar'' category, which includes many galaxies that, in our classification, would likely be labeled as disks with a clumpy feature flag, interacting/merger, or undefined when disk classification is uncertain. Another methodological difference lies in the choice of imaging filters: classified galaxies based on images taken in the filter that best matched the rest-frame optical wavelength at the source redshift (i.e., minimizing $\lambda_{\rm rest}/(1+z)$), whereas we utilized imaging across all available filters.

In addition to disk fractions, the classifications of galaxy morphologies also allow the measurement of the fraction of disks that appear featureless, i.e., lacking bars, bulges, clumps, or spiral arms. A recent study by \citet{Smethurst_2025} used Galaxy Zoo (\citealp{Lintott_2008}) classifications to analyze this population at $z \approx 3-7$, finding that although many galaxies are disk-like, most lack visible internal structure. To enable a direct comparison with our sample, we focused on their reported featureless disk fractions in the redshift range $z \approx 0.8-2.5$, which span fractions of $\approx 0.75-0.59$ with a mild decline toward lower redshift. After estimating a correction for observational constraints and redshift effects, they infer that the intrinsic featureless disk fractions are more likely in the range $\approx 0.38-0.29$. In our visual classifications, based on six expert classifiers and a different morphological scheme, we found featureless\footnote{Featureless refers to galaxies that were not classified as having bulges, clumps, bars, or spiral arms, to approximate the definition used by Smethurst et al., despite differences in classification schemes and criteria.} disk fractions of 0.47 and 0.20 in the high- and low-redshift bins, respectively. This redshift trend, opposite to that of \citet{Smethurst_2025}, may reflect differences in classification methodology, image depth, or sample selection: our study targets more massive galaxies at cosmic noon, whereas theirs includes a broader population over a wider redshift range. Nonetheless, the average values are remarkably similar (0.34 in our case vs. 0.32 from their bias-corrected estimates), suggesting that despite methodological differences, the underlying physical picture may be broadly consistent.

\subsection{Spirals}
\label{subsection: Spirals evolution}

We use the results from the visual classification of spiral galaxies to assess the spiral fraction in the two redshift bins, as $f_\mathrm{spiral} = N_\mathrm{spiral}/N_\mathrm{total}$. We found spiral fractions of 0.55 and 0.20 in the low- and high-redshift bins, respectively. It is worth noting that the remaining disks in our sample, which are not confidently classified as spirals, could still include galaxies with spiral features. These galaxies might not meet our criteria for confident identification due to factors such as near-edge-on orientations or small angular sizes. Consequently, the inferred spiral fractions could represent only a subset of the real population of disk galaxies with spiral features, as we prioritized robust classification over completeness.

As shown in Figure \ref{fig: Spiral fractions}, our inferred spiral fractions show general agreement with \cite{Kuhn_2024}, who also utilize \JWST imaging. We compare both their uncorrected and corrected values, where the latter account for observational effects, to estimate the intrinsic spiral fraction. While the overall trends are consistent, our results show a steeper decline in the spiral fraction with increasing redshift. As seen in \cite{Kuhn_2024}, applying corrections for these effects has a larger impact on the inferred intrinsic spiral fractions at higher redshifts. Both our results and those of Kuhn et al. find systematically higher spiral fractions compared to \cite{Margalef-Bentabol2022}, whose measurements are based on \HST imaging. This discrepancy is likely due to \textit{JWST}'s superior sensitivity and improved spatial resolution in the near-IR, which enables the identification of more subtle spiral structures that might be missed in \HST data. Additionally, \textit{JWST}’s coverage of longer rest-frame wavelengths enhances the visibility of the underlying disk, boosting confidence in the identification of spirals. 

This consistency supports the trend of a decreasing spiral fraction with increasing redshift, likely reflecting the progressive dynamical settling of disks over cosmic time. The presence of well-defined spirals at $z\approx 2$ suggests that some disks were already mature enough to sustain spiral structures, though they become increasingly rare at earlier epochs. However, this interpretation remains sensitive to observational limitations. As emphasized by \citet{Kuhn_2024}, the apparent decline is significantly reduced when accounting for these effects.

\begin{figure}
	\includegraphics[width=0.49\textwidth]{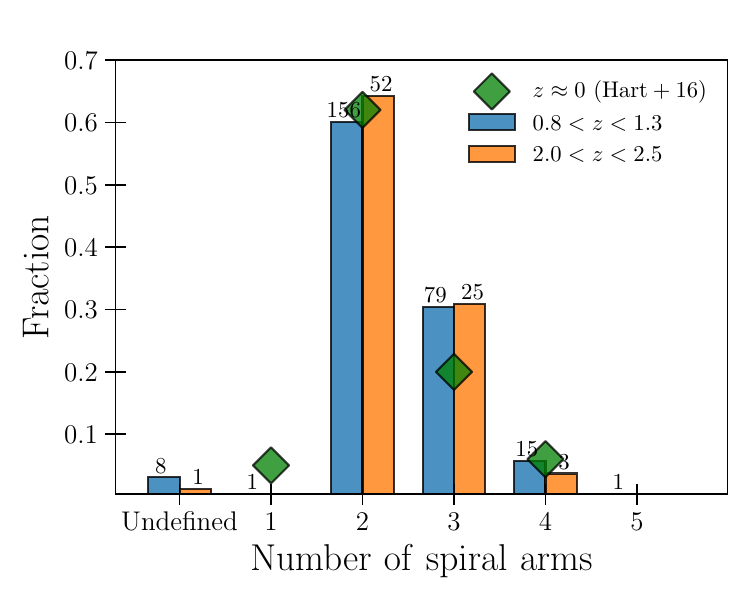}
	\caption{Distribution of the number of spiral arms identified from visual classification, using a threshold of four out of six classifiers for agreement. The undefined category refers to systems where classifiers agreed that no specific number of spiral arms could be discerned. The y-axis represents the fraction of spiral galaxies in each redshift bin, normalized by the total number of classified spiral galaxies in that bin. The blue bars correspond to galaxies in the $0.8 < z < 1.3$ range, while the orange bars represent galaxies in the $2.0 < z < 2.5$ range. Numbers displayed above the bars indicate the absolute counts of galaxies in each category. The green diamonds correspond to the debiased fractions from the luminosity-limited sample of \protect\cite{Hart_2016}.}
    \label{fig: Results of visual counting of arms}
\end{figure}

Beyond the overall fraction of spirals, their internal structure also shows striking consistency across redshifts. Applying the same 4/6 voting threshold as in the broad morphological classification, we determined the number of spiral arms in the subset of galaxies that met this criterion. Among the 574 classified galaxies, 341 surpassed this threshold, representing 60\% of the sample. This includes 260 out of 444 galaxies in the low-redshift bin (59\%) and 81 out of 130 in the high-redshift bin (62\%). We show the number of galaxies as a function of their identified number of spiral arms in Figure \ref{fig: Results of visual counting of arms}, where the percentages are normalized by the number of spiral galaxies in each redshift bin, for which a consensus of 4/6 votes was reached. We found relatively close agreement between the two redshift bins in the normalized arm number fractions: approximately two-thirds of galaxies exhibit two arms and about one-third show three arms in both bins. This suggests little evolution in spiral arm multiplicity across these cosmic epochs.

Interestingly, our visual classification of spiral arm multiplicity reveals trends broadly consistent with those observed in the local Universe. In both the $0.8 < z < 1.3$ and $2.0 < z < 2.5$ bins, two-armed spirals remain the dominant morphology, and have fractions comparable to that reported by \citet{Hart_2016}, who analyzed 62,220 Sloan Digital Sky Survey \citep[SDSS;][]{York_2000} galaxies as part of the Galaxy Zoo project in the redshift range $0.03 < z < 0.085$. For comparison, we used their luminosity-limited subsample ($M_r \leq -21$), which spans the same stellar mass range as our data ($\log(M_\star/M_\odot) = [10,11.5]$) and includes debiased fractions to correct for classification biases arising from redshift-dependent effects. Although we did not apply a formal debiasing correction, this adjusted low-redshift baseline provides a useful reference point for interpreting trends at higher redshift. 

The agreement in two-arm dominance suggests that grand-design spiral patterns were already common by $z \sim 1$–2 and supports the idea that many high-redshift disks were already cold and massive enough to sustain global $m=2$ modes, either spontaneously or aided by torques introduced by bars, or through external drivers such as interactions (e.g., \citealp{Toomre_1972, Athanassoula_1984, D'Onghia_2013, Dobbs_2014}). Although the prevalence of two-armed spirals appears stable, their structural properties may still evolve. For instance, \citet{Chugunov_2025} found that pitch angles tend to increase with redshift, suggesting that high-redshift spirals are more loosely wound than their low-redshift counterparts.

Beyond the dominant two-arm population, we found a noticeably higher incidence of three-armed spirals in our sample: while \citet{Hart_2016} reported that a fraction of about 0.20 of local spirals have three arms, we found a higher fraction ($\approx 0.30$), pointing to a modest increase in three-fold symmetry at earlier epochs. The relative excess of three-armed systems in our study could be the result of different disk properties at cosmic noon, such as higher gas fractions (e.g., \citealp{Tacconi_review_2020}). Interestingly, recent simulations have identified the emergence of three-armed spirals in turbulent, gas-rich disks at high redshift, preceding the formation of a two-armed spiral or a bar (\citealp{Bland-Hawthorn_2024}). Notably, one of the earliest high-redshift spirals observed with resolved structure also exhibits a three-armed morphology (\citealp{Law_2012}). In agreement with these results, our results offer the first evidence for an increased incidence of the three-armed spiral morphologies at cosmic noon.

\subsection{Bars}
\label{subsection: Bars evolution}

\begin{figure}
    \centering
    \includegraphics[width=0.5\textwidth]{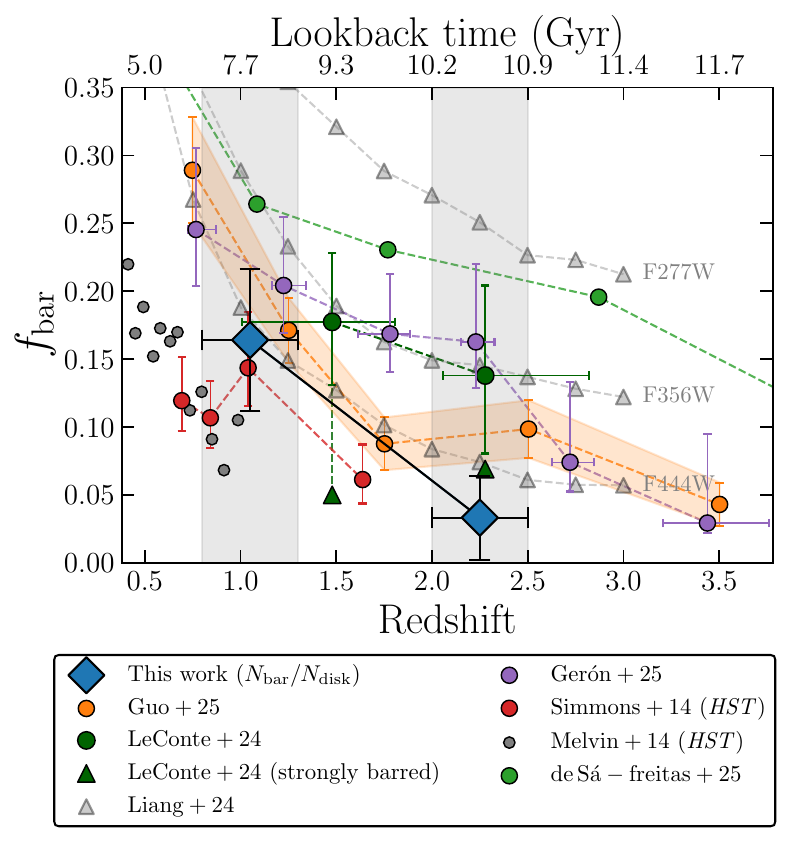}
    \caption{Bar fractions ($f_\mathrm{bar} = N_\mathrm{bar}/N_\mathrm{disk}$) from the subsample of disks with $e<0.5$ where the isophotal fitting analysis is reliable. The orange dots show the results from \protect\cite{Guo_2025}, while the green dots correspond to the fractions inferred by \protect\cite{LeConte_2024}, where the green triangles indicate the fractions inferred using only their sample of strongly barred galaxies. The purple dots indicate the Galaxy-Zoo based measurements from \protect\citep{Geron_2025} using \textit{JWST}-NIRCam imaging. The gray triangles represent simulated bar fractions from \protect\cite{Liang_2024} for a set of \JWST filters, given an intrinsic $f_\mathrm{bar}$ of 0.68 at $z\approx0$. Finally, the red and gray dots indicate the \textit{HST}-based bar fractions from \protect\citep{Simmons_2014} and \protect\citep{Melvin_2014}, respectively.}
    \label{fig: Bar fractions}
\end{figure}

We estimated the bar fractions as $f_\mathrm{bar} = N_\mathrm{bar}/N_\mathrm{disk}$, obtaining $\approx 0.16\pm0.05$ in the low-redshift bin and $\approx 0.03\pm0.03$ in the high-redshift bin. This overall decline with redshift is consistent with theoretical expectations that bars form more readily in dynamically cold, well-settled disks, which become increasingly common at lower redshifts. Supporting this, \citet{Sheth_2008} found that bar fractions exceed 0.60 at $z\approx 0$, but are expected to decline toward earlier epochs due to higher gas fractions and increased turbulence in young disks. Stellar population studies of local barred galaxies also support this physical picture: \citet{de-Safreitas_2025} inferred that many bars were already in place at high redshift, showing cumulative bar fractions at a given redshift that are broadly consistent with morphological studies, based on their backward extrapolations.

Further context comes from Galaxy Zoo classifications based on both \HST and \JWST imaging. Early studies using \HST data, including \citet{Melvin_2014} and \citet{Simmons_2014} reported bar fractions that already showed a decline in $f_\mathrm{bar}$ as a function of redshift up to $z\approx 1.7$. More recently, \citet{Geron_2025} extended Galaxy Zoo classifications to \JWST NIRCam imaging and found systematically higher bar fractions compared to the \textit{HST}-based studies at similar redshifts but a similar decreasing trend. The step up from \HST to \JWST, while still consistent with a declining trend, suggests that part of the observed evolution is driven by detection biases. The improved resolution and sensitivity of \JWST allow the identification of fainter and smaller bars that were missed in earlier classifications. As shown in Figure~\ref{fig: Bar fractions}, our results lie between the \HST and \JWST Galaxy Zoo measurements, closely matching the latter in the low-redshift bin, but falling below it in the high-redshift regime. While our visual classification approach differs from the Galaxy Zoo citizen science framework in both scope and methodology, the agreement at low redshift and divergence at higher redshift further highlight the influence of detection thresholds and classification systematics on $f_\mathrm{bar}$ estimates, effects that become increasingly significant at high redshift.

Our measured bar fractions align more closely with other \JWST studies that use expert classifications or ellipse fitting, such as \citet{Guo_2025} and \citet{LeConte_2024}. This likely reflects methodological similarities, in contrast to the Galaxy Zoo approach used by \citet{Geron_2025}, which adopts a different classification framework. Guo et al. used NIRCam F200W and F444W data from CEERS, combining ellipse fitting and visual inspection to identify bars. Le Conte et al., using CEERS+PRIMER data in F444W, reported higher bar fractions by including both strongly and weakly barred systems. In their classification scheme, galaxies with at least 3/5 votes for barred were labeled strongly barred, while those with 2/5 votes for barred or at least 3/5 for maybe-barred were considered weakly barred. When considering only the strongly barred subset, their bar fractions drop to 0.05–0.07 for the redshift range $1<z<3$, closely matching our measurements within the uncertainties.

Even with the improved resolution and sensitivity of \textit{JWST}, bar detectability remains strongly dependent on the choice of filter and its corresponding rest-frame wavelength coverage. \citet{Liang_2024} demonstrated this by artificially redshifting local galaxies and simulating \textit{JWST}/NIRCam observations across multiple filters, incorporating realistic size and luminosity evolution.  Adopting a conservative detection threshold of $a_\mathrm{bar} > 2 \times$ the PSF FWHM, they recovered a declining trend in bar fraction with redshift, but also found systematic offsets as a function of filter, reporting higher bar fractions in bluer bands and lower values in redder bands, as illustrated in Figure~\ref{fig: Bar fractions}.

Additional observational limitations must be carefully considered when interpreting the decline in bar fractions. Surface brightness dimming can obscure low-contrast bars, especially in faint or diffuse galaxies. \citet{Lian_2024} showed that near face-on spirals that have faint arms may be misclassified as edge-on systems, further complicating bar identification. In addition, spatial resolution constraints limit the detectability of small-scale bars, particularly at high redshift, where structural features approach the PSF scale and may remain unresolved. In our sample, for instance, bars smaller than 1.5–2 times the FWHM of the F444W PSF are not detected, consistent with the resolution threshold reported by \citet{Erwin_2018}. These limitations suggest that the observed decline in bar fractions with redshift may reflect not only intrinsic evolution but also increasing challenges in detection at earlier epochs. The true bar fraction at high redshift may be higher than what is currently measurable.

This possibility becomes even more compelling in light of recent observations of rapid radial gas inflows in high-redshift star-forming barred galaxies (\citealp{Genzel_2023}; \citealp{Arriagada-Neira_2025}; \citealp{Pastras_2025}) and even in dusty star-forming galaxies at $z>3$ (e.g., \citealt{Umehata_2025, Amvrosiadis_2025}. If such inflows are common in early disk galaxies, they could play a crucial role in shaping the evolution of the host galaxies, modifying the conditions for bar instabilities (e.g., \citealp{Bland-Hawthorn_2024}), potentially leading to an increase in the intrinsic bar fractions. Moreover, if baryons dominate the local gravitational potential over dark matter in these high-redshift galaxies \citep[e.g.,][]{Genzel_2017, Genzel_2020, Price_2021, Nestor_2023}, stellar bars could form on rapid timescales of 1–2 Gyr, even in gas-rich turbulent disks (e.g., \citealt{Fujii_2018, Bland-Hawthorn_2023, Bland-Hawthorn_2024}).

Complementary insights come from cosmological simulations, which generally predict a relatively constant intrinsic bar fraction across cosmic time. For example, \citet{Rosas-Guevara_2022}, using the TNG50 simulation suite (\citealp{Pillepich_2019}), find that bar fractions remain roughly flat at $\sim0.40$ from $z = 0$ to $z \approx 2$. Similarly, earlier work by \citet{Zhao_2020}, based on TNG100 \citep{Nelson_2019}, found an intrinsically constant, but observationally dropping bar fraction up to $z\sim1$. Independently, Frosst et al. (in prep.), using the COLIBRE cosmological hydrodynamical simulations (Schaye et al., in prep.), report a similarly flat bar fraction over this redshift range. Taken together, cosmological simulations and emerging observational evidence for bar-driven inflows at high redshift support the view that the observed decline in bar fractions is partly real but significantly amplified by detection limitations, leaving many high-redshift bars potentially hidden.

\subsection{Interacting and merger systems}
\label{subsection: Mergers evolution}

The fraction of galaxies classified as interacting or merging systems shows a clear decline with cosmic time, decreasing from $0.14\pm0.04$ in the higher redshift bin ($2.0 < z < 2.5$) to $0.06\pm0.02$ in the lower-redshift bin ($0.8 < z < 1.3$).  This trend aligns with the expected decrease in merger activity at lower redshifts, driven by declining galaxy densities and interaction rates (e.g., \citealp{Conselice_2007, Lin_2008, Lotz_2008, Rodriguez-Gomez_2015}). Despite our modest sample size, this decline remains consistent with theoretical and observational predictions in the framework of the hierarchical growth model in which major mergers become less frequent at later cosmic times.

\section{Quantitative morphological metrics}
\label{section: Morphological metrics}

We performed a quantitative analysis of the morphological properties of the sample to complement the visual classification. To achieve this, we employed several established nonparametric morphological indicators commonly utilized in morphological studies to characterize the structural features of the galaxies in our sample. These metrics include the concentration, asymmetry, and smoothness (CAS) statistics (\citealp{Conselice_2003}), the multinode, intensity, and deviation (MID) statistics (\citealp{Freeman_2013}), and the Gini and $M_{20}$ coefficients (\citealp{Lotz_2004}). All measurements are quantified using \statmorph (\citealp{Rodriguez_Gomez_2019}). Here is a brief description of them.

CAS: 
Concentration (C) quantifies the distribution of light within a galaxy, typically by comparing the radii containing 80\% and 20\% of its light, characterizing the galaxy's core. Asymmetry (A) evaluates the symmetry of the light distribution by comparing the original image with a version rotated by 180 degrees, often used to identify mergers or irregularities. Smoothness (S), also referred to as clumpiness, measures the fraction of light associated with small-scale structures, distinguishing smoother elliptical galaxies from clumpier, irregular galaxies or prominent spiral arms. 

\begin{figure}
	\includegraphics[width=0.495\textwidth]{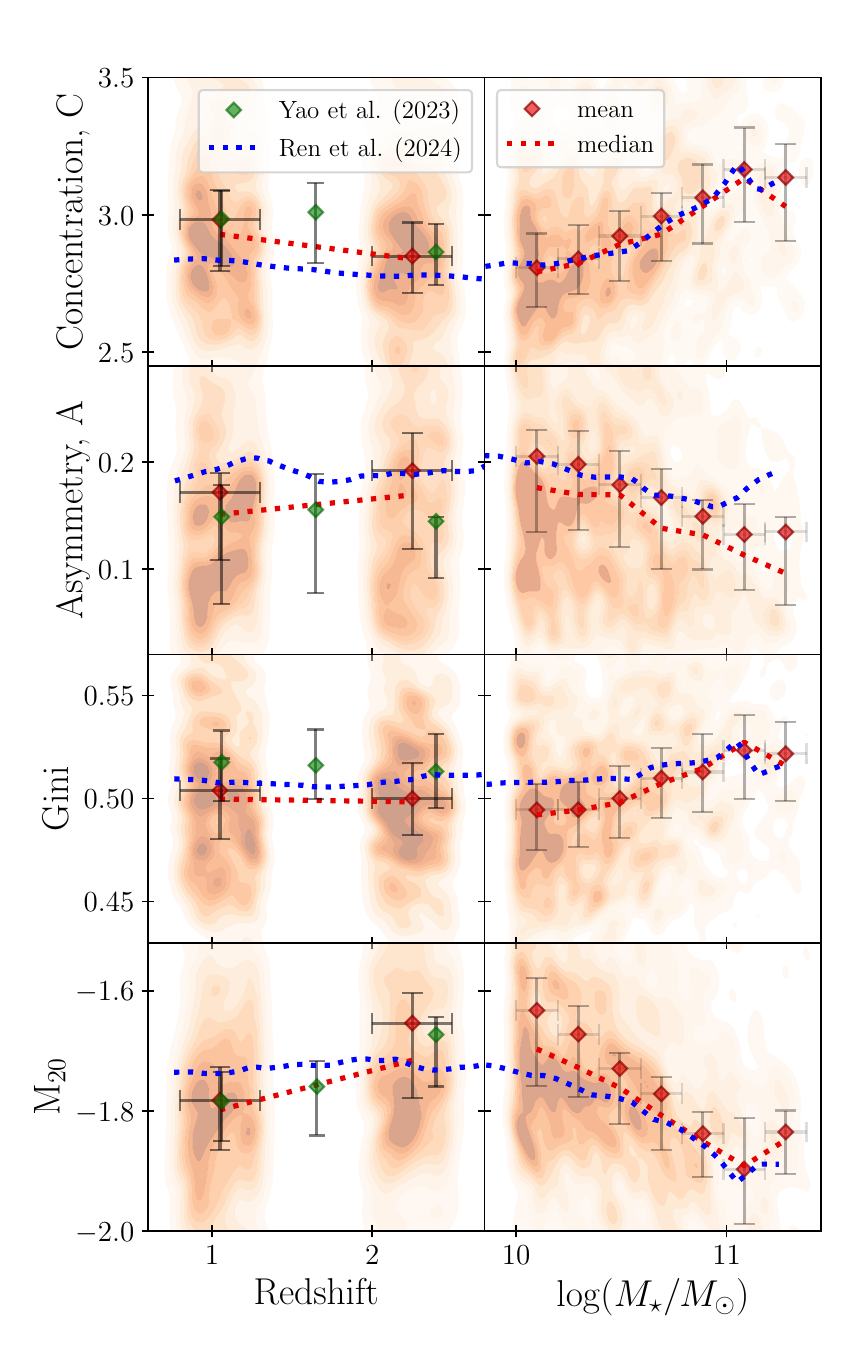}
    \centering
	\caption{Morphological statistics of galaxies across redshift (left column) and stellar mass (right column). From top to bottom: Concentration (C), asymmetry (A), Gini, and $M_{20}$ indices. Our results are presented as mean values (diamonds) and medians (horizontal markers). For comparison, we include the trends reported by \cite{Yao_2023} (green diamonds) over the range $0.8 < z < 3$ and \protect\cite{Ren_2024} (blue dotted lines) for $0.5 < z < 3$. Our error bars indicate the central 34\% range (spanning the 33rd to 67th percentiles) to ensure consistency with the results of Yao et al. The values from Ren et al. were originally adjusted for observational effects, such as noise and finite PSF resolution, using a correction formula. To ensure a consistent comparison with our F444W-matched galaxy sample, we reverse-applied their correction formula to de-adjust their values, aligning them with the observational conditions in our work and those by Yao et al. We find a strong agreement between our results and those of \protect\cite{Yao_2023} and \protect\cite{Ren_2024}, highlighting the consistency across studies.}
    \label{fig: Ren and Yao comparison}
\end{figure}

\begin{figure*}
    \includegraphics[width=0.99\linewidth]{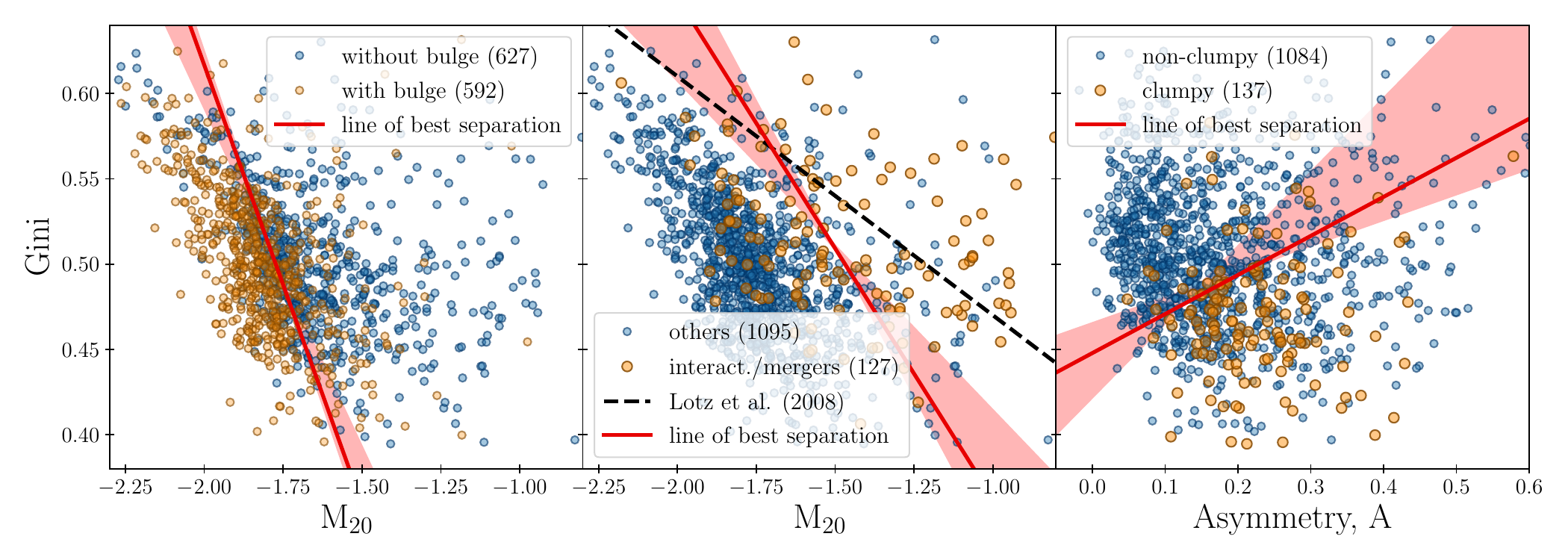}
    \caption{Notable parameter spaces where the combination of morphological metrics separate well different morphologies. In each panel, the red line corresponds to the line of best separation, which corresponds to the minimal contamination $\mathcal{C}_\mathrm{min}$ (defined in Appendix \ref{appendix: Separation Efficiency}), and the light red area indicates other possible line positions which achieve contamination within 0.01 of the minimal contamination, i.e., $C\in (\mathcal{C}_\mathrm{min},\mathcal{C}_\mathrm{min}+0.01)$. \textit{Left:} Gini \rm{vs} $M_{20}$ plane for galaxies with and without bulges. The line $G = -0.52 M_{20} - 0.4$, separates the main cluster of galaxies with bulges (71\%) and those without bulges (74\%). \textit{Middle:} Gini \rm{vs} $M_{20}$ plane for galaxies identified as interacting/mergers. The line $G = -0.29 M_{20} + 0.07$ (red) shows a systematic separation of the cluster of data points in this parameter space, with 50\% of visually classified mergers and 93\% of nonmergers correctly separated. For comparison, we show with the black dashed line the boundary line established by \protect\cite{Lotz_2008} and discuss the difference in the main text. \textit{Right:} Gini \rm{vs} asymmetry A plane for clumpy and nonclumpy galaxies. The separation line is given by $G = 0.229 A+0.448$, where 81.8\% of clumpy galaxies are below the line and 64.1\% of the nonclumpy galaxies are above.}
    \label{fig: Gini vs M20 all}
\end{figure*}

MID:
The multinode, intensity, and deviation (MID) statistics allow the identification of disturbed morphologies, such as mergers or interacting systems. Multinode (M) measures the presence of multiple peaks in the light distribution, indicating complex or multi-component structures. Intensity (I) assesses the brightness contrast between these peaks and their surroundings, highlighting areas of intense star formation or disruptions. Deviation (D) quantifies the departure of the light distribution from a smooth, symmetric model, further aiding in the detection of structural disturbances.

Gini and $\mathit{M_{20}}$ coefficients:  The Gini ($G$) coefficient characterizes the distribution of the flux among the pixels assigned to an extended source. It varies from 0 to 1, where $G=0$ corresponds to a uniform distribution (all pixels contain the same flux), and $G=1$ corresponds to the flux being concentrated in a single pixel. The $M_{20}$ coefficient, which quantifies the second-order moment of the brightest 20\% of the galaxy’s flux relative to the total light distribution, helps in identifying the spatial extent and configuration of bright regions, so it is a useful metric to quantify the degree of morphological disturbance.

We ensured consistency across the sample at different redshifts by applying PSF matching, selecting uniform rest-frame wavelengths, and masking with segmentation maps. The rest-frame wavelength for morphology analysis was set to $\lambda_\mathrm{RF}=1.37 \mu m$, chosen as it corresponds to the optimal wavelength where the majority of galaxies in both redshift bins in our sample (1,300 out of 1,451) have imaging, F277W for the low-redshift bin and F444W for the high-redshift bin, both of which are wide filters, ensuring high signal-to-noise ratios. Spatial resolution differences were minimized by matching the PSF to a target resolution corresponding to the largest physical PSF in the sample (1.19 kpc). The detailed methodology, including wavelength and PSF matching, segmentation map creation, and masking procedures, is described in Appendix \ref{appendix: Morphological metrics preparation}.

To assess the robustness of the morphological indicators, we conducted a systematic visual inspection of diagnostic images produced by \textsc{starmorph} (e.g., Figure 4 in \protect\citealp{Rodriguez_Gomez_2019}). The reliability of each measurement was assessed based on the following criteria: How well the model replicates the galaxy’s light distribution (e.g., the agreement between the observed and model images); quality of the residual image, where well-fit galaxies exhibit minimally structured residuals and poor fits show significant leftover flux or systematic artefacts; convergence of fitted parameters, ensuring they fall within physically reasonable ranges; and absence of contamination from neighbouring sources or image artefacts that could bias measurements. The evaluation of these diagnostic images, along with assessments of segmentation maps and PSF matching, were used to ensure the accuracy of the nonparametric morphological measurements. After applying this filtering process, we identified a high-quality subset of 1,222 galaxies that met our reliability criteria. We adopted this subset for the results and analysis presented in this work.

\subsection{Dependence on redshift and stellar mass}
\label{subsection: Dependence on redshift and stellar mass}

In the majority of the derived morphological indicators, including concentration (C), Gini, asymmetry, and $M_{20}$, we observe little to no evolution with redshift but notable trends with galaxy mass. Specifically, more massive galaxies tend to exhibit higher concentrations and Gini values, indicative of more compact and centrally concentrated light distributions, while $M_{20}$, indicative of morphological disturbance, shows a clear decreasing trend with increasing galaxy mass. The lack of significant redshift evolution suggests that the overall structural properties of galaxies, as captured by these metrics, remain relatively stable for $\sim 2.7$ Gyr between the median redshifts of 1.05 and 2.25 probed in our sample, and are primarily dependent on the stellar mass. These findings are consistent with trends previously identified in \HST studies (e.g., \citealp{Wuyts_2011, Lang_2014}).

\begin{figure}
    \centering
	\includegraphics[width=0.47\textwidth]{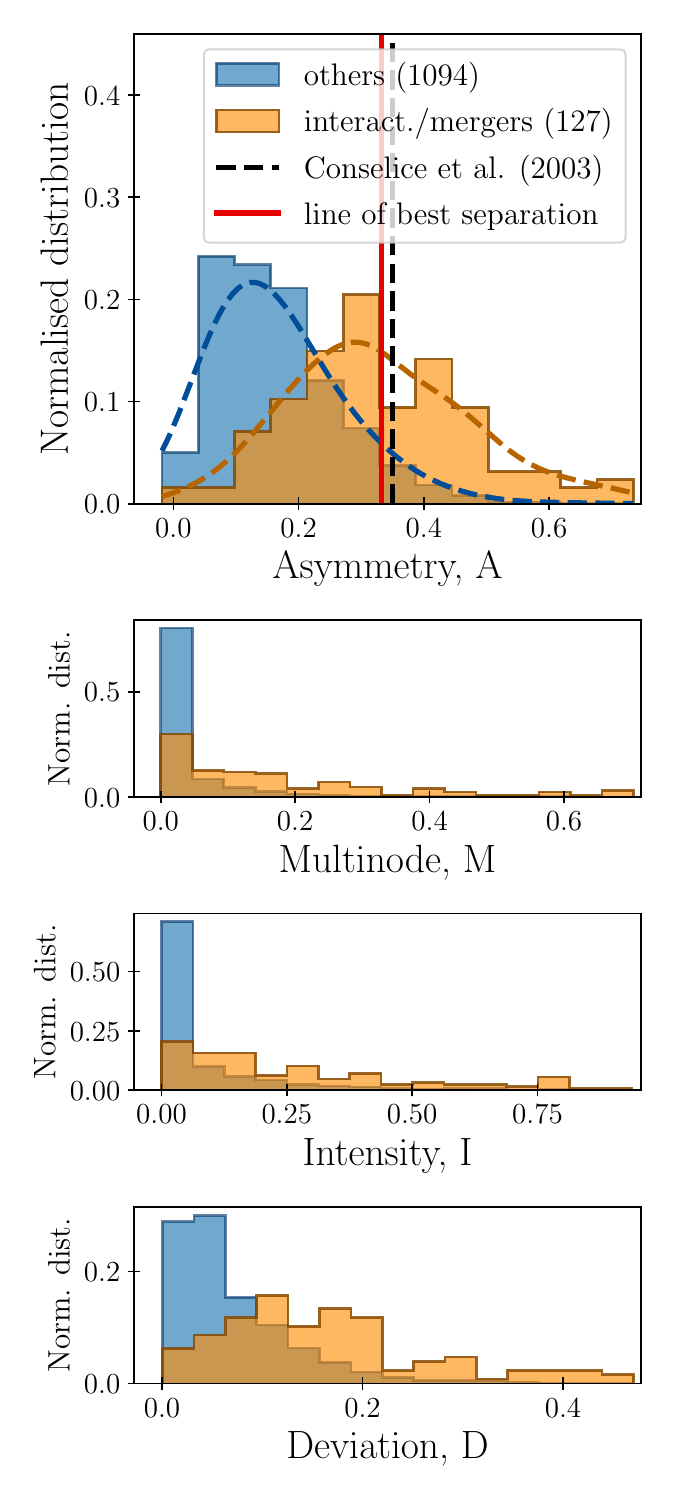}
    \caption{\textit{Top panel}: Distribution of the asymmetry statistics (A in CAS) in our sample. The statistics show a clear systematic separation around value A$=0.23$. Maintaining a merger fraction of $\approx10\%$, we obtained the vertical red line A$=0.332$ that separates 44\% of mergers and 93\% of nonmergers, with only a minor offset from the boundary line of A$=0.35$ introduced by \protect\cite{Conselice_2003}. \textit{Bottom three panels}: Multinode, intensity, and deviation (MID) statistics. They reveal systematic differences in distributions between visually classified mergers and nonmergers, as expected, given their design to capture morphological features of disturbed systems.}
    \label{fig: MID and CAS for mergers}
\end{figure}

Our findings also agree with trends reported in the literature, particularly in \cite{Yao_2023} and \cite{Ren_2024}, which similarly noted stable morphological indicators across redshifts but a clear dependence on galaxy stellar mass. Figure \ref{fig: Ren and Yao comparison} illustrates these comparisons, showing the mean values reported by \cite{Yao_2023} for a F444W-matched sample in the redshift range $0.8 < z < 3$, as well as the median values from \cite{Ren_2024} for $0.5 < z < 3$. To enable a consistent comparison, we reverse-applied the observational corrections used in \cite{Ren_2024} to their data, which accounted for effects such as noise and finite PSF resolution, aligning their values with our F444W-matched dataset. In both cases, we found a strong agreement between the trends in our analysis and those reported by these studies. 

\subsection{Connection with visual classification}
\label{subsection: Connection with visual classification}

To use the extensive set of morphological parameters computed in this analysis, we systematically explored how different galaxy classifications separate across various parameter spaces derived from \statmorph. Specifically, we identified parameter combinations that effectively distinguish galaxies with and without bulges, mergers and nonmergers, and clumpy versus nonclumpy galaxies. To quantitatively assess the separation between classifications, we computed a ``separation efficiency'' metric based on misclassification rates. The methodology for defining and calculating this metric, including the optimization of classification boundaries, is detailed in Appendix \ref{appendix: Separation Efficiency}, where we provide a full set of separation efficiencies for each classification. Given the wealth of possible parameter combinations, we highlight only the most informative and physically meaningful ones, prioritizing those that offer the clearest distinctions between morphological classes that have also been studied in the literature. However, all possible combinations, including those not well studied in other studies, are presented in Tables \ref{tab: bulges separation}-\ref{tab: clumpy separation}.

We emphasize that widely used morphological metrics, originally designed and quantified from \HST imaging, must be carefully reassessed using \textit{JWST}'s improved resolution, wavelength coverage, and depth. However, in this study, we focus on identifying broad, top-level trends in our sample rather than proposing precise recalibrations of these classification boundaries. A robust recalibration, which is beyond the scope of this work, should be carried out using larger statistical samples that fully leverage the growing wealth of high-quality \JWST data, as well as the extensive galaxy catalogs from ongoing large surveys produced by Euclid, which will provide millions of galaxies at local and intermediate redshifts (e.g., \citealt{Huertas-Company_2025_EUCLID, Walmsley_2025_EUCLID}).

Bulges: For galaxies visually identified as containing bulges, the Gini–$M_{20}$ plane provides one of the strongest separations between bulge-dominated and nonbulge galaxies (left panel of Figure \ref{fig: Gini vs M20 all}). The optimal separation line, given by $G = -0.52 M_{20} - 0.41$, correctly separates 74\% of galaxies without bulges, as shown by the clear systematic offset between the two populations in the figure. This offset aligns with previous studies (e.g., \citealp{Conselice_2003}; \citealp{Lotz_2008}), which demonstrated that nonparametric morphology indicators are effective in distinguishing bulge-dominated galaxies. The correlation between bulge presence, high Gini values, and low $M_{20}$ values supports the interpretation that bulge-dominated galaxies have centrally concentrated light distributions and less fragmented substructures, making them easier to identify in this parameter space. Specifically, low $M_{20}$ values indicate that a galaxy’s brightest pixels are concentrated near the center, which is characteristic of systems that contain prominent bulges.  

Beyond the Gini–$M_{20}$ plane, $M_{20}$ alone also serves as a strong discriminator. Additionally, combining $M_{20}$ with either the \sersic index or the effective radius further improves classification accuracy. A full summary of bulge separation efficiency is presented in Table \ref{tab: bulges separation} in the Appendix.

Mergers: Several parameter combinations effectively distinguish interacting and merging galaxies, including the Gini–$M_{20}$ plane, MID statistics, and CAS parameters. The full separation efficiencies for mergers and nonmergers are presented in Table \ref{tab: mergers separation} in the Appendix.  

In the Gini–$M_{20}$ plane, we define a separation line that maintains a merger fraction of $\approx 10 \%$, matching both our measured fraction and the findings of \cite{Lotz_2008}. The resulting threshold, given by $G = -0.29 M_{20} + 0.07$, correctly captures 93\% of nonmergers (middle panel of Figure \ref{fig: Gini vs M20 all}). For comparison, we show the boundary line estimated by a dedicated study of mergers and morphology \cite{Lotz_2008}, which analyzed $\approx3,000$ galaxies with \HST imaging at $0.2<z<1.2$ and found a separation boundary of $G = -0.14 M_{20} + 0.33$. We found notable differences between our classification and theirs, particularly a 0.15 slope difference in the separation line. Moreover, our separation achieves a contamination value of 0.57, while adopting the Lotz et al. boundary in our dataset would result in a higher contamination value of 0.74. The differences in our boundary lines likely stem from differences in sample selection, imaging depth, rest-frame wavelength, spatial resolution, and classification objectives. Whereas Lotz et al. optimized their boundary specifically for merger detection, we focused on a wider range of morphologies.  Despite these differences, the overall separation trends remain similar, reinforcing that Gini vs $M_{20}$ remains a useful discriminator and is effective, at least in a global sense, within our sample.

\begin{figure*}
\centering
	\includegraphics[width=0.99\textwidth]{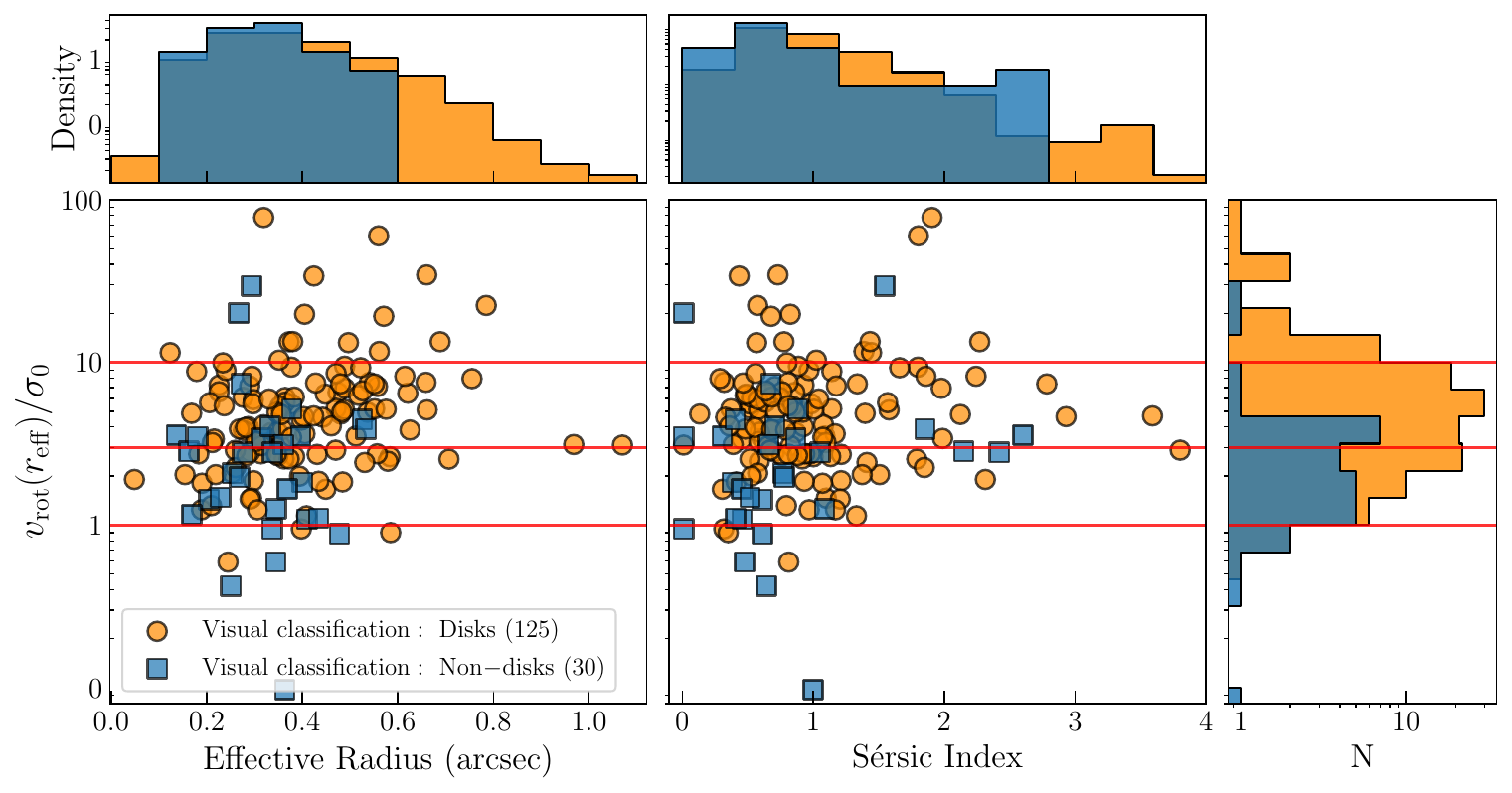}
	\caption{Scatter plots of $v_{\mathrm{rot}}(r_{\mathrm{eff}})/\sigma_0$ as a function of effective radius (left) and \sersic index (right) for disk (orange circles) and nondisk (blue squares) galaxies. The histograms above each panel show the normalized distributions of effective radius and \sersic index, while the histogram on the right represents the distribution of $v_{\mathrm{rot}}(r_{\mathrm{eff}})/\sigma_0$. The red horizontal lines indicate reference values at $v_{\mathrm{rot}}(r_{\mathrm{eff}})/\sigma_0 = 1, 3,$ and $10$.}
    \label{fig: v over sigma sersic}
\end{figure*}

Among the CAS statistics, the asymmetry parameter provides another useful merger discriminator. Figure \ref{fig: MID and CAS for mergers} (top panel) illustrates the separation, where visually identified mergers are skewed toward higher asymmetry values, while nonmergers peak at lower asymmetries. The separation threshold that maintains a merger fraction of 10\% closely aligns with the empirical merger criterion proposed by \cite{Conselice_2003} (dashed black), reinforcing the utility of asymmetry as a merger diagnostic.

Similarly, the MID statistics also effectively separate mergers from nonmergers. As shown in the bottom panels of Figure \ref{fig: MID and CAS for mergers}, the multinode (M), intensity (I), and deviation (D) distributions exhibit systematic differences between the two populations. Interacting systems consistently show higher values across all three statistics, reinforcing the idea that MID metrics capture morphological disturbances and asymmetries associated with mergers and their irregular light distributions.

Clumps: The Gini coefficient is particularly effective in distinguishing clumpy from nonclumpy galaxies, especially when combined with other morphological metrics. Among these, the Gini–A (asymmetry) plane provides the clearest separation (right panel of Figure \ref{fig: Gini vs M20 all}), highlighting the strong correlation between light concentration and structural irregularity in clumpy systems. The complete set of separation efficiencies for clumpy galaxies is provided in Table \ref{tab: clumpy separation} in the Appendix.

\section{Morpho-kinematic Correlations}
\label{section: Kinematics}

For this discussion, we focus on global morpho-kinematic trends on a population-wide angle rather than detailed substructures, examining how the identified morphologies relate to dominant dynamical support as measured by $v/\sigma$. We used the homogenized $v_\mathrm{rot}(\re)$ and $\sigma_0$ values obtained from various subsamples outlined in Sect. \ref{subsection: Kinematics sample}. Our analysis reveals a notable distinction in the dynamical properties of galaxies classified as disks versus nondisks. For the 155 galaxies that have kinematics measurements, 125 (fraction $\approx 0.81$) were visually classified as disks, while 30 ($\approx 0.19$) were not identified as disks.

The histogram on the right of Figure \ref{fig: v over sigma sersic} presents the $v/\sigma$ distributions for visually classified disks and nondisks. A high $v/\sigma$ ratio suggests a dynamically cold, rotation-dominated system typical of disks, while lower values indicate dynamically hot, dispersion-dominated systems more consistent with spheroidal geometries or disturbed morphologies. We include horizontal lines at $v/\sigma = 1$, 3, and 10—commonly adopted thresholds in the literature corresponding to varying degrees of rotational support (e.g., \citealp{Genzel_2006}; \citealp{Kassin_disk_settling}; \citealp{KMOS3D}; \citealp{Wisnioski_2019}; \citealp{Stott_2016}; \citealp{Tiley}; \citealp{Gillman}; \citealp{Forster_review_2020}).

To quantify the difference between the two populations, we performed a Kolmogorov–Smirnov (KS) test, which yields a statistic of 0.39 and a $p$-value of $9.4 \times 10^{-4}$, confirming that the $v/\sigma$ distributions of disks and nondisks are significantly different. Specifically, among the 125 visually classified disk galaxies, 88 (fraction $\approx0.70$) exhibit $v/\sigma > 3$, indicating strong rotational support. The remaining 37 ($\approx 0.30$) fall below this threshold, suggesting either weaker rotational support or the presence of noncircular motions (pure turbulence but also radial motions contributing to higher $\sigma_0$ in modeling of the typically seeing-limited IFU data included here). For nondisk galaxies, the trend is reversed: only 12 galaxies ($\approx 0.40$) have $v/\sigma > 3$, while the majority, 18 galaxies ($\approx 0.60$), have $v/\sigma < 3$. This distribution reflects the expectation that nondisk galaxies exhibit lower rotational support, as their dynamics are often dominated by random motions rather than ordered rotation. In the case of marginal rotational support of $v/\sigma=1$, only three galaxies identified as disks and five from the other categories are below this boundary.

The apparent correlation between disk morphology and high $v/\sigma$ is reassuring, yet it is important to consider observational limitations in the kinematic measurements that could affect such correlation. One such effect is beam smearing, which artificially enhances velocity dispersions in smaller galaxies (e.g., \citealp{Newman_2013}). In particular, compact galaxies at high redshift may exhibit inflated velocity dispersions due to unresolved velocity gradients, leading to an underestimation of their rotational support. Other key observational limitations include spectral resolution and the signal-to-noise ratio (S/N) of the emission lines (e.g., \citealp{Davies_2011}). At high S/N, velocity dispersions can be reliably measured down to one-third of the spectral resolution, but at low S/N, noise-driven line broadening inflates dispersion estimates, biasing $v/\sigma$ to lower values (e.g., \citealp{KMOS3D, Uebler_2019}). 

Given the impact of beam smearing and S/N limitations, the observed correlation between disk morphology and high $v/\sigma$ might be even stronger than what we found, since some dynamically cold disks could be misclassified due to observational effects.

We further investigated how effective radius and \sersic index correlate with measured $v/\sigma$ ratios. To do this, we used the best-fit \sersic parameters derived with \textsc{statmorph} on the sample that was matched in spatial resolution and rest-frame wavelength, as discussed in Sect. \ref{section: Morphological metrics}. The scatter plots in Figure \ref{fig: v over sigma sersic} show $v/\sigma$ ratios as a function of $\re$ and $n$. In terms of the \sersic indices, we found no discernible trend since there is a clear clustering of both disks and nondisks below a threshold of $n = 1.5$. In terms of the galaxy size, we found a weak yet statistically significant positive correlation for disks between $v/\sigma$ and the effective radius, with a Spearman correlation coefficient ($\rho = 0.26$, $p = 0.004$). While the correlation is weak, it aligns with theoretical expectations and the broader connection between galaxy size and dynamical structure observed in both simulations and observational studies (e.g., \citealp{Dekel, Fall_Romanowsky_2013, Obreschkow_2014, Teklu, Posti_2018, El-Badry2018a, Tiley_2021, Espejo_Salcedo_2022, Espejo_Salcedo_2025}). Furthermore, larger disks are easier to visually identify in our classifications, making their structural properties more straightforward to analyze, as discussed in Sect. \ref{subsection: Classification overview} and shown in Figure \ref{fig: Disk fraction vs effective radius}.

Beyond the technical aspects of classification agreement, these results provide insight into the nature of galaxy evolution at these redshifts. The observed connection between disk morphology and high $v/\sigma$ suggests that a significant fraction of star-forming galaxies at cosmic noon are already dynamically settled and rotation-dominated. This is consistent with findings that a large fraction of galaxies at $z\sim1-3$ exhibit regular rotation (e.g., \citealp{KMOS3D, Johnson, Genzel_2020, Tiley_2021, Nestor_2023}). Moreover, we also found that a substantial fraction of these rotationally supported systems exhibit non-axisymmetric structures. Of the kinematically analyzed disks, 80 (fraction 0.64) are classified as spirals, and 12 (0.10) are classified as barred, ten of which also have spiral structure. The median $v/\sigma$ values further support the idea that these structures form in dynamically cold disks: spirals show a median $v/\sigma \approx 7$ and bars show a slightly lower value of $v/\sigma \approx 5$. This modest drop in rotational support for barred galaxies may reflect added kinematic complexity associated with bar-driven secular evolution, in line with recent discussions on angular momentum redistribution and dynamical stability in barred disk galaxies (e.g., \citealp{Romeo_2023}).

\section{Conclusions}
\label{section: Conclusions}

In this study, we analyzed a selected sample of \JWST NIRCam images to investigate the morphological properties of 1,451 massive star-forming galaxies ($\log (\mathrm{SFR}/\mathrm{SFR_{MS}}) = [-0.4,0.4]$) within a mass range of $\log(M_\star/M_{\odot}) = [10.0, 11.4]$, and sizes within $\re \, \mathrm{(arcsec)} = [0.2,1.0]$ at two key epochs within within cosmic $0.8<z<1.3$ and $2.0<z<2.5$. We performed a global visual classification and identified key substructures such as bars and spirals. We complemented this with a quantitative analysis of morphology as a function of redshift and stellar mass and investigated links between morphology and kinematic properties where available. Our main findings are the following:

\begin{itemize}
    \item Global morphologies: Based on our visual classifications, we found that the majority ($0.82\pm0.03$) of our massive main-sequence galaxies exhibit disk morphology, while smaller fractions are spheroidal ($0.02\pm0.01$), interacting/mergers ($0.10\pm0.02$), and undefined ($0.02\pm0.01$). We found that $0.05\pm0.02$ of the galaxies have companions, and we only identified three lensed systems ($<0.01$). The rest of the sample did not reach the 4/6 voting threshold of our classification scheme.
    
    \item Spirals: A fraction of $0.48\pm0.04$ of visually identified disks feature spiral structures, and show a decline from $0.55\pm0.05$ in the low-redshift bin to $0.20\pm0.05$ in the high-redshift bin. A follow-up classification assessed the number of spiral arms, revealing similar distributions across both redshift bins: approximately two-thirds of spirals have two arms (fractions of $0.60\pm0.09$ and $0.64\pm0.16$), and about one-third have three arms ($0.30\pm9$ and $0.31\pm15$). Four-armed spirals remain rare ($0.06\pm0.04$ and $0.04\pm0.06$). While the two-armed spiral fraction is consistent with results at $z = 0$ ($\approx0.60$), we report for the first time an increased incidence of three-armed spirals from $z\approx 0$ to cosmic noon, rising from $\approx0.20$ to $\approx0.30$.

    \item Bars: We identified 84 barred systems using isophotal fitting, where 74 are in the low-redshift bin and ten are in the high-redshift bin, corresponding to bar fractions among the disk galaxies of $0.16\pm0.05$ and $0.03\pm0.03$, respectively. These values align with previous studies on bar identification. While the decline in bar fraction as a function of redshift may reflect intrinsic evolution, it is also influenced by observational challenges, such as surface brightness dimming, resolution limitations, and the expected evolution of galaxy and bar sizes. Overcoming these limitations requires greater sensitivity to surface brightness and improved spatial resolution, particularly in the infrared capabilities that future observatories such as the ELT will provide.
    
    \item Quantitative morphological metrics: Nonparametric indicators indicate no significant evolution with redshift but a strong dependence on mass. Exploring all possible parameter combinations, we identified optimal metrics for distinguishing specific morphological features, consistent with previous studies: \textit{i)} Bulges are best separated by Gini-$M_{20}$ and by $M_{20}$ alone. \textit{ii)} Mergers are well distinguished using Gini-$M_{20}$, MID statistics, and CAS parameters. \textit{iii)} Clumps are best identified with the Gini coefficient alone and the combination of Gini-asymmetry.

    \item Kinematics: We found a correlation between disk morphology and high rotational support. A fraction of $\approx 0.70$ of kinematically analyzed disks meet a $v/\sigma > 3$ criterion. Disks with high $ v/\sigma $ also tend to have \sersic indices near unity and correlate modestly with $\re$, reinforcing the reliability of morphology-based classifications. A significant fraction of these rotationally supported galaxies exhibit non-axisymmetric structures: $\approx 0.64$ are spirals, and $\approx0.10$ are barred (all these barred systems are spirals). Additionally, both spirals and bars show large $v/\sigma$ values: spirals have a median $v/\sigma \approx 7$, while bars have $v/\sigma \approx 5$. While this difference is modest, the slightly lower $v/\sigma$ in barred galaxies may reflect additional kinematic effects, such as radial motions increasing velocity dispersions. 

\end{itemize}

Building on the findings presented here, we plan several follow-up studies. These include an investigation of the barred subsample and their kinematics to understand bar structures at high redshift better (Pulsoni et al., in prep), and an analysis of the connection between rapid inflows and non-axisymmetric substructures in gas-rich systems (Pastras et al., in prep). We also intend to examine metallicity gradients in disk galaxies to assess the role of radial mixing driven by bars and spiral arms.

\begin{acknowledgements}
      We thank the anonymous referee for the constructive and helpful feedback and suggestions to improve the quality of this paper. We also thank Pascale Monier and Helenka Kinnan for their valuable suggestions on the language, layout, and overall presentation of the manuscript. We further thank Maciejewski, A. Marchuk, S. Arriagada-Neira, and A. Burkert, for their valuable discussions and suggestions during the development of this work. J.M.E.S., N.M.F.S, G.T., C.B., and J.C. acknowledge financial support from the European Research Council (ERC) Advanced Grant under the European Union’s Horizon Europe research and innovation programme (grant agreement AdG GALPHYS, No. 101055023). HÜ acknowledges funding by the European Union (ERC APEX, 101164796). Views and opinions expressed are however those of the authors only and do not necessarily reflect those of the European Union or the European Research Council Executive Agency. Neither the European Union nor the granting authority can be held responsible for them. R.H.-C. thanks the Max Planck Society for support under the Partner Group project "The Baryon Cycle in Galaxies" between the Max Planck for Extraterrestrial Physics and the Universidad de Concepción. R.H-C. also gratefully acknowledge financial support from ANID - MILENIO - NCN2024\_112 and ANID BASAL FB210003. This work is based on observations made with the NASA/ESA/CSA \JWST and complementary \HST data obtained from the Mikulski Archive for Space Telescopes (MAST) at the Space Telescope Science Institute (STScI) which is operated by the Association of Universities for Research in Astronomy, Inc., under NASA contract NAS 5-03127 for \textit{JWST}, and NAS 5-26555 for \textit{HST}. The observations can be accessed via \href{https://doi.org/10.17909/hyn4-m554}{DOI:10.17909/hyn4-m554}.. We acknowledge the DAWN \JWST Archive at the Cosmic Dawn Centre funded by the Danish National Research Foundation under grant DNRF140. We further appreciate the open-source software packages used throughout this work, including \textsc{astropy} (\citealp{astropy}), \textsc{scipy} (\citealp{Virtanen_2020_scipy}), \textsc{numpy} (\citealp{numpy}), CM\textsc{asher} (\citealp{cmasher}), \textsc{matplotlib} (\citealp{matplotlib}), \galfit (\citealp{Peng_2002, Peng_2010}), \statmorph (\citealp{Rodriguez_Gomez_2019}), \textsc{trilogy} \citep{Coe_2012}.
\end{acknowledgements}

\bibliographystyle{aa}
\bibliography{references}

\begin{appendix} 
\onecolumn

\label{appendix: Appendix}

\section{Observational limitations}
\label{appendix - Observational limitations}

\begin{itemize}
    \item Spatial resolution: The spatial resolution of \JWST NIRCam is set by its PSF; the FWHM ranges from 0.035 to 0.16 arcseconds depending on the filter. This corresponds to 0.28-1.29 kpc at $z=0.8-1.3$ and 0.29-1.32 kpc at $z=2.0-2.5$. Given the near-identical spatial scales across the two bins, resolution differences are unlikely to introduce systematic biases in the morphological classifications between them.
    
    \item Cosmological dimming: Cosmological surface brightness dimming, which scales as $(1+z)^4$, causes high-redshift galaxies to appear substantially fainter per unit area. Between our two redshift bins, where the mean values are $z \approx 1.05$ and $z \approx 2.25$, this corresponds to a $\sim6\times$ reduction in surface brightness. This effect primarily suppresses extended, low-surface-brightness components such as disks and spiral arms, while compact, bright structures such as bulges and clumps remain visible. As a result, classifications at higher redshift may be biased toward irregular morphologies and undercount disk or spiral features, even in deep \JWST imaging.
    
    \item Wavelength: The appearance of galaxy morphology is sensitive to the observed wavelength, as different wavelengths trace distinct stellar populations and structural components. Redder filters trace older stellar populations, are less affected by dust extinction, and capture large-scale structures such as bulges and disks, but may miss finer features such as narrow spiral arms or short bars, biasing classifications against subtle substructures. Such biases translate to systematic variations in the inferred S\'ersic parameters.  To assess how this impacts our analysis, we select the 743 galaxies (51\% of the sample) with imaging in all eight common NIRCam filters (F090W, F115W, F150W, F200W, F277W, F356W, F410M, and F444W). We compute the median S\'ersic parameters in each band and find clear trends: $n$ increases and $\re$ decreases with wavelength (Fig. \ref{fig: Sersic parameters vs wavelength}). While our morphological classifications leverage all filters jointly, such residual wavelength-dependent biases may persist due to implicit weighting from signal-to-noise and visibility in individual bands.
\end{itemize}

\begin{figure}[h!]
\centering
	\includegraphics[width=0.58\textwidth]{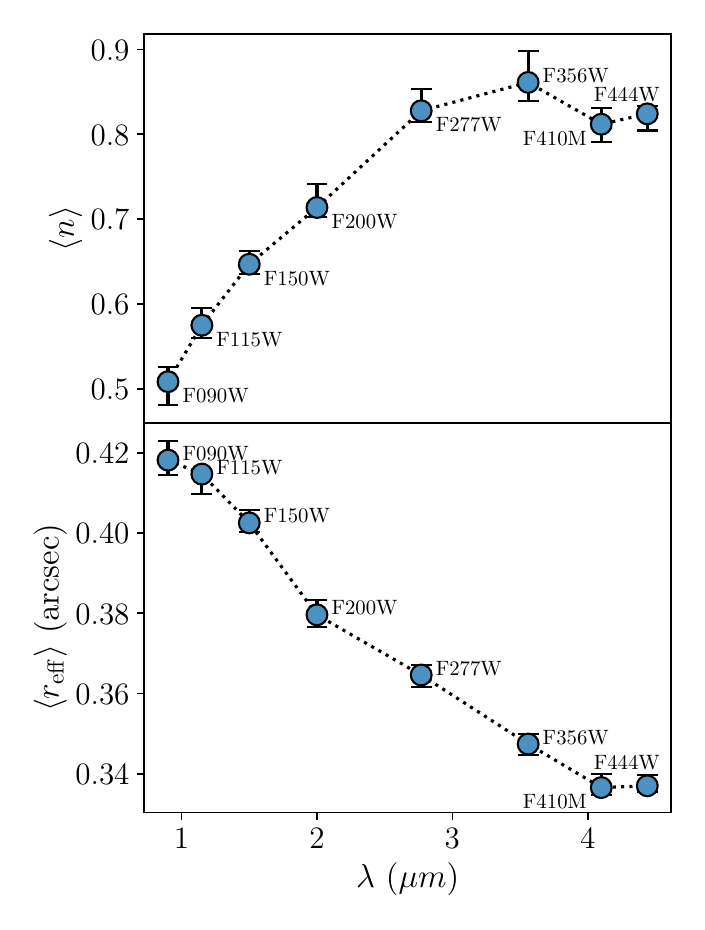}
	\caption{S\'{e}rsic parameters \rm{vs} wavelength for the 739 galaxies with the largest combination of filters, which corresponds to 51\% of the full sample. The values correspond to the median values in the corresponding wide-band filters with 68\% confidence interval error bars, computed using bootstrapped resampling.}
    \label{fig: Sersic parameters vs  wavelength}
\end{figure}

\section{Morphological metrics preparation}
\label{appendix: Morphological metrics preparation}

We describe below the preparatory steps before running \textsc{statmorph}, used to homogenize the dataset.

\subsection{Wavelength and PSF matching}
\label{subsection: Wavelength and PSF matching}

The quantification of morphological metrics using unmodified \JWST data is expected to show redshift-dependent variations, driven both by intrinsic evolution in galaxy structure and by observational effects. These observational effects are primarily due to spatial resolution, cosmological dimming, and wavelength dependence, as discussed above.

\begin{figure}[h!]
\centering
	\includegraphics[width=0.7\textwidth]{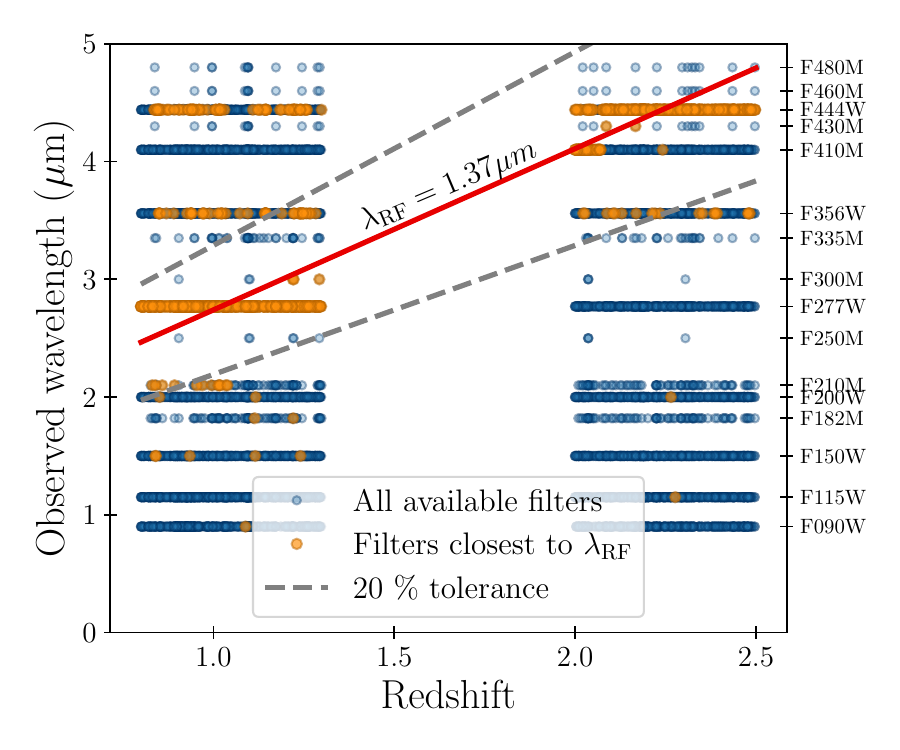}
	\caption{Optimal rest-frame wavelength determined to be $\lambda_\mathrm{RF} = 1.37 \mu$m (red solid line). Approximately 1,300 galaxies in the sample have imaging in filters corresponding to this value, within a 20\% margin of error ($\pm 0.27 \mu m$) indicated by the dashed lines. Most galaxies in the low-redshift bin are observed in the F277W filter, while those in the high-redshift bin are predominantly observed in the F444W filter. The plot illustrates the wavelengths of the available filters (blue) and the closest matching filter to $\lambda_\mathrm{RF}$ for each galaxy at its respective redshift (orange).}
    \label{fig: optimal rfw}
\end{figure}

To distinguish between intrinsic evolution and observational effects, we modified the data to create a uniform sample, ensuring fair comparisons across galaxies. To minimize rest-frame wavelength differences, we selected the observed wavelength closest to the target rest-frame wavelength ($\lambda_\mathrm{RF} = 1.37\mu$m) for each galaxy, discarding cases where no suitable match within $\pm 20\%$ was available, as illustrated in Figure \ref{fig: optimal rfw}.

\begin{figure}[h!]
\centering
	\includegraphics[width=0.92\textwidth]{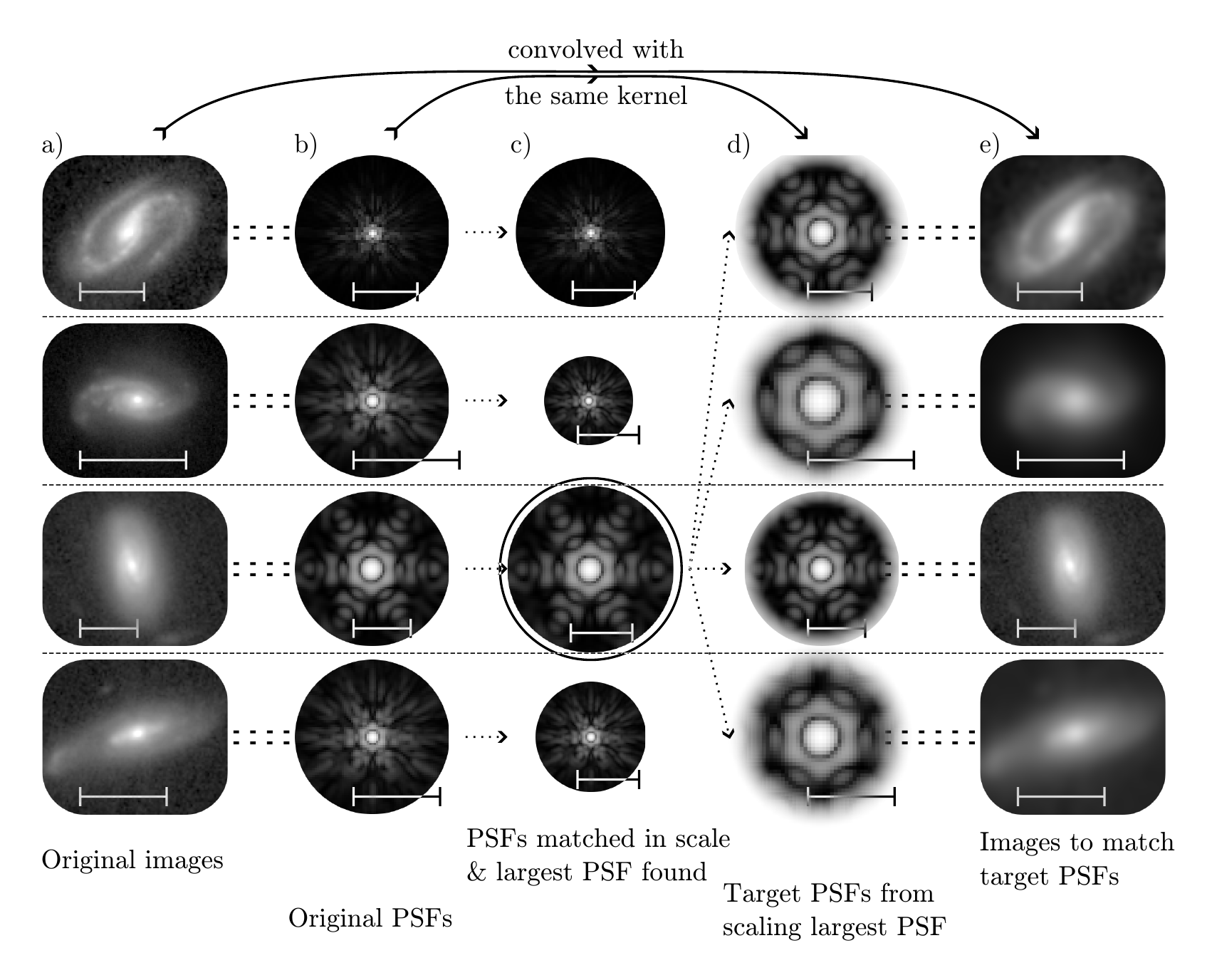}
	\caption{Illustration of the method used to match the spatial resolution of galaxy images across the sample, so that all galaxies appear as they would under identical observing conditions. (a) Original images: Initial galaxy images, with a 2 kpc horizontal scale bar indicating their physical size at the galaxy’s redshift. (b) Original PSFs: The point spread functions (PSFs) corresponding to each image. (c) Scaling the PSFs: To enable a fair comparison, PSFs are rescaled to represent physical sizes (kpc) consistently. The FWHM of each PSF is measured, and the galaxy with the largest PSF is identified. In this case, the largest PSF corresponds to a galaxy at $z=2$ observed with the F444W filter, with a PSF width of 1.19 kpc. (d) Creating target PSFs: The largest PSF from (c) is used as a reference. It is rescaled to match each galaxy's physical size while maintaining a constant FWHM in spatial units. This ensures that all PSFs correspond to the same physical resolution across the sample. This is evident in the main PSF component, which spans a fixed fraction (approximately half) of the horizontal scale bar across all images. (e) Final image matching: Each original image (from panel a) is convolved with a kernel derived from the rescaled PSFs (panel d). This convolution results in a final set of images that all share the same spatial resolution, making them directly comparable.}
    \label{fig: PSF matching}
\end{figure}

To address spatial resolution differences, we convolved the images to a common point spread function (PSF), as shown in Figure \ref{fig: PSF matching}. The steps for making the sample uniform are outlined below:

\begin{enumerate}[label=\textit{\roman*)}] 
    \item Filter Selection: Each galaxy is assigned its corresponding filter based on the matched rest-frame wavelength ($\lambda_\mathrm{RF} = 1.37 \mu m$), along with its corresponding PSF. 
    \item Spatial Resolution Calculation: The spatial resolution for each galaxy is determined by converting the width of its PSF (e.g., FWHM) into physical units (e.g., kpc) at the galaxy's redshift. 
    \item Target PSF Identification: The galaxy with the lowest spatial resolution (largest physical PSF size) is identified. For our sample, this corresponds to observing an object at $z\approx2$ with the F444W filter, equivalent to a PSF width in physical units of $1.19$ kpc. This becomes the target PSF to which the entire sample is matched.
    \item PSF Rescaling: For each galaxy, the target PSF is rescaled to match the pixel scale of that galaxy's image.
    \item Kernel Creation: A convolution kernel is generated to transform the PSF of each galaxy to the rescaled target PSF. 
    \item PSF Matching: The data and PSF of each galaxy are convolved with the kernel to match the spatial resolution to the target PSF. The accuracy of this process is assessed by comparing the resulting PSF to the scaled target PSF. 
\end{enumerate}

By accounting for observational effects, the modifications reduce the resolution of the data, resulting in some loss of information. However, this trade-off is essential for minimizing biases toward the lower-redshift objects (e.g., overestimating clumpiness). Additionally, for \statmorph calculations, the adjustments improve reliability by mitigating issues in segmentation and masking, particularly in lower-wavelength images where such calculations are prone to fail. Despite these benefits, it is acknowledged that some previously resolved features may be lost due to the decreased resolution.

\subsection{Segmentation map and masking}
\label{subsection: Segmentation map and masking}
 
\begin{figure}[h!]
\centering
	\includegraphics[width=0.99\textwidth]{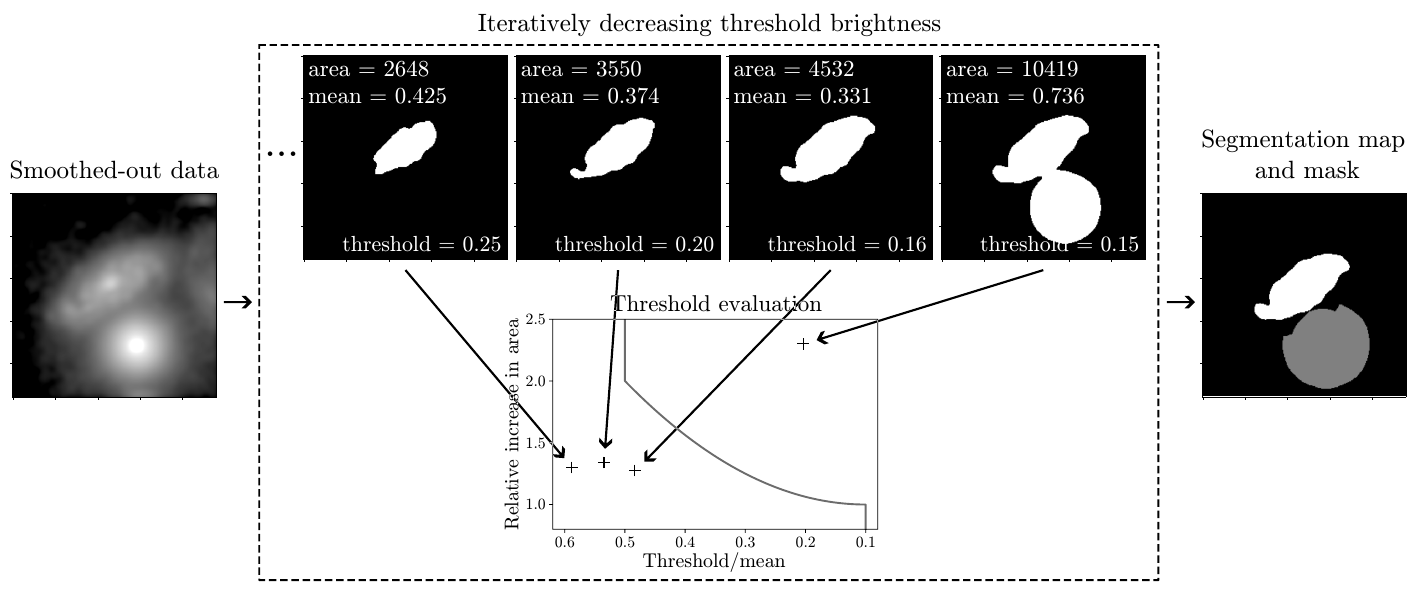}
	\caption{Illustration of the iterative process used to determine the brightness threshold for constructing the segmentation map. We start by smoothing out the image data with a 10-pixel-wide Gaussian. Then, at each step, the brightness threshold is decreased incrementally, and the impact is assessed by examining (1) the relative increase in the area enclosed by the threshold (y-axis) and (2) the ratio of the threshold value to the mean brightness within the enclosed area (x-axis). The threshold is deemed sufficiently low when the target object can be distinguished from other objects in the image, which is determined by whether the cutoff curve (gray) has been crossed in this parameter space. The optimal shape of the cutoff curve was established through manual evaluation of a randomly selected subset of the sample.}
    \label{fig: segmap threshold}
\end{figure}

The precision of morphological indicator calculations was enhanced by applying masks generated from segmentation maps with brightness thresholds. These thresholds were determined on an image-by-image basis using a custom iterative method. The process involved incrementally lowering the threshold while tracking changes in two parameters: the ratio of the threshold to the mean brightness of the central object and the relative increase in the enclosed area. An approximate outline of this method is illustrated in Figure \ref{fig: segmap threshold}.

\section{Separation Efficiency in Morphological Parameter Spaces}
\label{appendix: Separation Efficiency}

For each pair of morphological indicators provided by \textsc{statmorph}, we evaluate how effectively mergers can be distinguished from nonmergers, galaxies with bulges from those without bulges, and clumpy galaxies from nonclumpy galaxies within their respective parameter spaces. This is done by iteratively determining an optimal linear boundary that minimizes contamination on both sides. We define contamination as the total fraction of misclassified objects across all categories. It is computed as the sum of misclassification rates, where one corresponds to the fraction of objects that belong to a category but were not classified as such (false negatives rate), and the other one is the fraction of objects incorrectly assigned to that category (false positives rate). The summed misclassification rates provide a simple measure of the overall contamination. Table \ref{tab: Contamination definition} illustrates the parameters used in this calculation, where the contamination $\mathcal{C}$ is defined as $\mathcal{C} = a/(a+b)+ c/(c+d)$.


\begin{table}[h!]
    \centering
    \caption{Definition of parameters used in the contamination calculations.}
    \begin{tabular}{ccc}
       \hline\hline 
       \diagbox{Classified}{Reference}  & In category & Not in category\\\hline
       In category      & $b$ & $a$\\
       Not in category  & $c$ & $d$
    \end{tabular}
\tablefoot{Rows indicate the visual category assigned to each object, while columns refer to the reference category, i.e., the morphological category against which contamination is evaluated. The entries denote the number of galaxies in each combination: $a$ and $c$ are misclassifications, where $a$ represents objects incorrectly classified as belonging to the given category, and $c$ represents objects that should have been classified as such but were not. The letters $b$ and $d$ are correct classifications. Contamination is computed as the sum of misclassification rates, as defined in Equation C.1.}
    \label{tab: Contamination definition}
\end{table}

The separation efficiency of a given parameter space is directly linked to contamination, defined as $1 - \mathcal{C}_{\mathrm{min}}$, where $\mathcal{C}_{\mathrm{min}}$ is the lowest contamination achievable within that space by any dividing line. The full separation efficiency results for galaxies with and without bulges, mergers and nonmergers, and clumpy and nonclumpy galaxies are presented in Tables \ref{tab: bulges separation}, \ref{tab: mergers separation}, \ref{tab: clumpy separation}, respectively. As an illustrative example, in the case of clumpy galaxies and for the Gini vs A combination, the minimal contamination corresponds to $\mathcal{C}_\mathrm{min} = \frac{25}{25+112}+\frac{389}{389+695} = 0.54$, and thus the separation efficiency corresponds to $1-\mathcal{C}_\mathrm{min} = 0.46$. The line of best separation that optimizes this separation efficiency is shown in the right panel of Figure \ref{fig: Gini vs M20 all}.

\begin{table*}[h!]
      \caption{Separation efficiency of galaxies with and without bulges across parameter spaces defined by various morphological indicators.}
    \label{tab: bulges separation}
    \centering
    \begin{tabular}{ccccccccccc}
    \hline \hline & Gini & $\text{M}_{20}$ & C & A & S & M & I & D & $n$ & $r_{\text{eff}}$\\\hline
    Gini & \cellcolor{red!0} 0.034 & \cellcolor{red!68} 0.453 & \cellcolor{red!45} 0.311 & \cellcolor{red!13} 0.119 & \cellcolor{red!29} 0.213 & \cellcolor{red!20} 0.157 & \cellcolor{red!24} 0.185 & \cellcolor{red!36} 0.255 & \cellcolor{red!34} 0.242 & \cellcolor{red!31} 0.223 \\
    $\text{M}_{20}$ &  & \cellcolor{red!58} 0.388 & \cellcolor{red!63} 0.422 & \cellcolor{red!63} 0.422 & \cellcolor{red!59} 0.397 & \cellcolor{red!59} 0.397 & \cellcolor{red!59} 0.393 & \cellcolor{red!62} 0.412 & \cellcolor{red!70} 0.459 & \cellcolor{red!69} 0.456 \\
    C &  &  & \cellcolor{red!23} 0.177 & \cellcolor{red!23} 0.179 & \cellcolor{red!33} 0.238 & \cellcolor{red!29} 0.213 & \cellcolor{red!25} 0.189 & \cellcolor{red!39} 0.275 & \cellcolor{red!28} 0.208 & \cellcolor{red!39} 0.275 \\
    A &  &  &  & \cellcolor{red!12} 0.111 & \cellcolor{red!30} 0.219 & \cellcolor{red!25} 0.192 & \cellcolor{red!21} 0.165 & \cellcolor{red!45} 0.311 & \cellcolor{red!26} 0.194 & \cellcolor{red!45} 0.312 \\
    S &  &  &  &  & \cellcolor{red!28} 0.208 & \cellcolor{red!33} 0.238 & \cellcolor{red!32} 0.232 & \cellcolor{red!44} 0.305 & \cellcolor{red!33} 0.240 & \cellcolor{red!44} 0.302 \\
    M &  &  &  &  &  & \cellcolor{red!19} 0.153 & \cellcolor{red!24} 0.185 & \cellcolor{red!36} 0.258 & \cellcolor{red!28} 0.205 & \cellcolor{red!52} 0.353 \\
    I &  &  &  &  &  &  & \cellcolor{red!19} 0.153 & \cellcolor{red!37} 0.260 & \cellcolor{red!25} 0.192 & \cellcolor{red!56} 0.379 \\
    D &  &  &  &  &  &  &  & \cellcolor{red!34} 0.246 & \cellcolor{red!39} 0.275 & \cellcolor{red!54} 0.366 \\
    $n$ &  &  &  &  &  &  &  &  & \cellcolor{red!25} 0.191 & \cellcolor{red!38} 0.265 \\
    $r_{\text{eff}}$ &  &  &  &  &  &  &  &  &  & \cellcolor{red!28} 0.205\\\hline
    \end{tabular}
    \tablefoot{The values represent the effectiveness of each parameter combination, where higher values indicate better separation. The combination of Gini and $M_{20}$ provides a prominent separation of bulges and nonbulges, consistent with its frequent use in morphological studies. $M_{20}$ alone also demonstrates strong discriminatory power. Additionally, the combination of $\re$ and intensity-based statistics performs comparably to the Gini and $M_{20}$ indicators.}
\end{table*}

\begin{table*}
    \caption{Similar to Table \ref{tab: bulges separation}, but for mergers and nonmergers.}
    \label{tab: mergers separation}
    \centering
    \begin{tabular}{ccccccccccc}
    \hline \hline & Gini & $\text{M}_{20}$ & C & A & S & M & I & D & $n$ & $r_{\text{eff}}$\\\hline
    Gini & \cellcolor{red!0} 0.132 & \cellcolor{red!63} 0.530 & \cellcolor{red!24} 0.288 & \cellcolor{red!69} 0.570 & \cellcolor{red!17} 0.244 & \cellcolor{red!58} 0.500 & \cellcolor{red!61} 0.521 & \cellcolor{red!61} 0.517 & \cellcolor{red!43} 0.405 & \cellcolor{red!11} 0.206 \\
    $\text{M}_{20}$ &  & \cellcolor{red!49} 0.445 & \cellcolor{red!65} 0.541 & \cellcolor{red!67} 0.553 & \cellcolor{red!50} 0.450 & \cellcolor{red!64} 0.536 & \cellcolor{red!64} 0.534 & \cellcolor{red!59} 0.508 & \cellcolor{red!54} 0.473 & \cellcolor{red!53} 0.466 \\
    C &  &  & \cellcolor{red!4} 0.160 & \cellcolor{red!69} 0.566 & \cellcolor{red!12} 0.212 & \cellcolor{red!58} 0.500 & \cellcolor{red!60} 0.515 & \cellcolor{red!63} 0.529 & \cellcolor{red!21} 0.265 & \cellcolor{red!5} 0.169 \\
    A &  &  &  & \cellcolor{red!66} 0.549 & \cellcolor{red!66} 0.549 & \cellcolor{red!70} 0.571 & \cellcolor{red!66} 0.549 & \cellcolor{red!68} 0.563 & \cellcolor{red!66} 0.549 & \cellcolor{red!66} 0.549 \\
    S &  &  &  &  & \cellcolor{red!10} 0.200 & \cellcolor{red!59} 0.508 & \cellcolor{red!60} 0.513 & \cellcolor{red!61} 0.515 & \cellcolor{red!23} 0.281 & \cellcolor{red!16} 0.233 \\
    M &  &  &  &  &  & \cellcolor{red!58} 0.500 & \cellcolor{red!63} 0.533 & \cellcolor{red!65} 0.545 & \cellcolor{red!59} 0.506 & \cellcolor{red!61} 0.521 \\
    I &  &  &  &  &  &  & \cellcolor{red!60} 0.510 & \cellcolor{red!67} 0.557 & \cellcolor{red!61} 0.517 & \cellcolor{red!62} 0.523 \\
    D &  &  &  &  &  &  &  & \cellcolor{red!59} 0.505 & \cellcolor{red!61} 0.518 & \cellcolor{red!59} 0.508 \\
    $n$ &  &  &  &  &  &  &  &  & \cellcolor{red!20} 0.261 & \cellcolor{red!20} 0.261 \\
    $r_{\text{eff}}$ &  &  &  &  &  &  &  &  &  & \cellcolor{red!4} 0.160\\\hline
    \end{tabular}

\end{table*}

\begin{table*}
    \caption{Similar to Table \ref{tab: bulges separation}, but for clumpy and nonclumpy galaxies.}
    \label{tab: clumpy separation}
    \centering
    \begin{tabular}{ccccccccccc}
    \hline \hline & Gini & $\text{M}_{20}$ & C & A & S & M & I & D & $n$ & $r_{\text{eff}}$\\\hline
    Gini & \cellcolor{red!58} 0.403 & \cellcolor{red!64} 0.431 & \cellcolor{red!64} 0.431 & \cellcolor{red!70} 0.459 & \cellcolor{red!64} 0.433 & \cellcolor{red!60} 0.409 & \cellcolor{red!60} 0.413 & \cellcolor{red!64} 0.432 & \cellcolor{red!60} 0.411 & \cellcolor{red!62} 0.420 \\
    $\text{M}_{20}$ &  & \cellcolor{red!4} 0.123 & \cellcolor{red!41} 0.315 & \cellcolor{red!53} 0.375 & \cellcolor{red!12} 0.164 & \cellcolor{red!29} 0.252 & \cellcolor{red!37} 0.292 & \cellcolor{red!7} 0.141 & \cellcolor{red!41} 0.314 & \cellcolor{red!46} 0.338 \\
    C &  &  & \cellcolor{red!31} 0.261 & \cellcolor{red!58} 0.399 & \cellcolor{red!34} 0.278 & \cellcolor{red!38} 0.300 & \cellcolor{red!49} 0.354 & \cellcolor{red!33} 0.272 & \cellcolor{red!40} 0.309 & \cellcolor{red!56} 0.389 \\
    A &  &  &  & \cellcolor{red!49} 0.355 & \cellcolor{red!49} 0.356 & \cellcolor{red!51} 0.364 & \cellcolor{red!50} 0.359 & \cellcolor{red!56} 0.387 & \cellcolor{red!52} 0.369 & \cellcolor{red!51} 0.366 \\
    S &  &  &  &  & \cellcolor{red!4} 0.126 & \cellcolor{red!29} 0.252 & \cellcolor{red!33} 0.275 & \cellcolor{red!9} 0.147 & \cellcolor{red!40} 0.310 & \cellcolor{red!43} 0.326 \\
    M &  &  &  &  &  & \cellcolor{red!29} 0.252 & \cellcolor{red!38} 0.299 & \cellcolor{red!30} 0.255 & \cellcolor{red!39} 0.303 & \cellcolor{red!44} 0.329 \\
    I &  &  &  &  &  &  & \cellcolor{red!33} 0.271 & \cellcolor{red!41} 0.312 & \cellcolor{red!47} 0.343 & \cellcolor{red!43} 0.323 \\
    D &  &  &  &  &  &  &  & \cellcolor{red!0} 0.101 & \cellcolor{red!46} 0.340 & \cellcolor{red!46} 0.338 \\
    $n$ &  &  &  &  &  &  &  &  & \cellcolor{red!39} 0.301 & \cellcolor{red!55} 0.387 \\
    $r_{\text{eff}}$ &  &  &  &  &  &  &  &  &  & \cellcolor{red!43} 0.322\\\hline
    \end{tabular}

\end{table*}

\end{appendix}

\end{document}